\documentclass[fleqn,usenatbib]{mnras}
\usepackage[T1]{fontenc}
\usepackage{ae,aecompl}
\usepackage{newtxtext,newtxmath}
\usepackage{graphicx}	
\usepackage{amsmath}	
\usepackage{color,ulem}
\usepackage{array}
\usepackage{tabularx}
\usepackage{tablefootnote}
\usepackage{subfigure}

\usepackage{color,ulem}  
\definecolor{deepblue}{rgb}{0,0,0.5}  
\definecolor{deepred}{rgb}{0.6,0,0}   
\definecolor{deepgreen}{rgb}{0,0.5,0} 
\definecolor{darkgreen}{rgb}{0,0.6,0} 

\def\eddr{$\lambda_{\rm Edd}$}

\def\ha{H$\alpha$}

\def\hanc{H$\alpha_{\rm NC}$}

\def\havbc{{\sc{H}}$\alpha_{\rm VBC}$\/}
\def\hb{H$\beta$}

\def\hbnc{H$\beta_{\rm NC}$}
\def\hbvbc{{\sc{H}}$\beta_{\rm VBC}$\/}
\def\hbsb{{\sc{H}}$\beta_{\rm SC}$\/}
\def\heii{{He\sc{II}}$\lambda$4686\/}
\def\hebc{{He\sc{II}}$_{\rm BC}$\/}
\def\henc{{He\sc{II}}$_{\rm NC}$\/}
\def\hg{H$\gamma$}

\def\hgnc{H$\gamma_{\rm NC}$}
\def\hgsb{{\sc{H}}$\gamma_{\rm SC}$\/}
\def\hgvbc{{\sc{H}}$\gamma_{\rm VBC}$\/}
\def\oi{{\sc [Oi]}}
\def\oi{{[O\sc{i}]}\/}
\def\oil{{\sc [Oi]}$\lambda$6300\/}

\def\oia{{[O\sc{i}]}$\lambda$6300\/}

\def\oiii{{[O\sc{iii}]}\/}

\def\oiiib{{[O\sc{iii}]}$\lambda$5007\/}
\def\oiiill{{[O\sc{iii}]}$\lambda\lambda$4959,5007\/}

\def\oiiig{{[O\sc{iii}]}$\lambda$4364\/}

\def\nii{{[N\sc{ii}]}\/}

\def\niib{{[N\sc{ii}]}$\lambda$6583\/}

\def\sii{{[S\sc{ii}]}$\lambda\lambda$6716,6731\/}
\def\siia{{[S\sc{ii}]}$\lambda$6716\/}
\def\siib{{[S\sc{ii}]}$\lambda$6731\/}

\def\l{$\lambda$}

\def\kms{km s$^{-1}$\/}

\def\mbh{M$\rm_{BH}$}

\def\lbol{L$_{\rm bol}$\/}
\def\ledd{L$_{\rm Edd}$}
\def\l5100{L$_{\it 5100}$}
\def\iras{IRAS\,05589+2828}
\def\twomas{2MASX\,J06021107+2828382}
\def\chandra{{\it Chandra}}
\def\swift{{\it Swift}}
\def\vla{{\it VLA}}

\title[Unravelling the nature of the dual AGN in the galaxy pair system]
{Unravelling the nature of the dual AGN in the galaxy pair system IRAS\,05589+2828 and 2MASX J06021107+2828382}

\author[E.~Ben\'{\i}tez et al.]
{E.~Ben\'{\i}tez,$^{1}$\thanks{Email: erika@astro.unam.mx}
E.~Jim\'enez-Bail\'on,$^{2}$
C.~A.~Negrete,$^{3}$
D.~Ruschel-Dutra,$^{4}$
\newauthor
J.~M.~Rodr\'iguez-Espinosa,$^{5,6}$
I.~Cruz-Gonz\'alez,$^{1}$
L.~F.~Rodr\'iguez,$^{7}$
\newauthor
V.~H.~Chavushyan,${^8}$
P.~Marziani,$^{9}$
L.~Guti\'errez,$^{2}$
O.~Gonz\'alez-Martin,$^{7}$ 
\newauthor
B. W.~Jiang$^{10,11}$ and
M. D'Onofrio$^{12}$
\\
\\
$^{1}$Universidad Nacional Aut\'onoma de M\'exico, Instituto de Astronom\'\i{}a, AP 70-264, CDMX  04510, Mexico \\
$^{2}$Universidad Nacional Aut\'onoma de M\'exico, Instituto de Astronom\'\i{}a, AP 106, Ensenada 22800, BC, Mexico\\
$^{3}$CONACYT Research Fellow -Universidad Nacional Autónoma de México, Instituto de Astronom\'\i{}a, AP 70-264, CDMX  04510, Mexico\\
$^{4}$Departamento de F\'isica, Universidade Federal de Santa Catarina, P.O. Box 476, 88040-900, Florian\'opolis, SC, Brazil\\ 
$^{5}$Instituto de Astrof\'isica de Canarias, V\'ia L\'actea, s/n, 38205, La Laguna, Spain\\
$^{6}$Departamento de Astrof\'isica, Universidad de La Laguna (ULL), 38205, Spain\\
$^{7}$Instituto de Radioastronom\'\i{}a y Astrof\'\i{}sica, Universidad Nacional Aut\'onoma de M\'exico, AP 3-72, (Xangari), 58089, Morelia, Mexico\\
$^{8}$ Instituto Nacional de Astrof\'isica, \'Optica y Electr\'onica, Luis Enrique Erro \#1, Tonantzintla, Puebla 72840, Mexico\\
$^{9}$Istituto Nazionale d'Astrofisica (INAF), Osservatorio Astronomico di Padova, 35122, Padova, Italy\\
$^{10}$Key Laboratory for Particle Astrophysics, Institute of High Energy Physics, Chinese Academy of Sciences, 
19B Yuquan Road, Beijing 100049, China\\
$^{11}$School of Astronomy and Space Science, University of Chinese Academy of Sciences, 19A Yuquan Road, Beijing 100049, China\\
$^{12}$Dipartimento di Fisica e Astronomia, Universit\`a di Padova, 35122, Padova, Italy\\
}

\date{Accepted August 05. Received August 05; in original form February 07}

\pubyear{2022}

\begin{document}
\label{firstpage}
\pagerange{\pageref{firstpage}--\pageref{lastpage}}

\maketitle

\begin{abstract}
We have studied the nuclear region of the previously detected dual AGN system in the galaxy pair IRAS 05589+2828 and 2MASX J06021107+2828382 through new optical spectroscopy observations, along with radio and X-ray archival data. Our multiwavelength data strongly suggest that the Sy1 \iras\, (z=0.0330$\pm$0.0002) conforms to a dual AGN system with the Sy2 \twomas\, (z=0.0334$\pm$0.0001) with a projected separation obtained from the radio data of 20.08\arcsec\, ($\sim$13.3\,kpc). Analysis of the optical spectra reveals a faint narrow extended emission from H$\alpha$ and [OIII] amidst the two AGN, supporting evidence for an ongoing merger. \iras\, is a double component narrow emission line AGN, with complex broad Balmer emission line profiles that clearly show a strong red-peaklet with a velocity shift of $\sim$3500\,km\,s$^{-1}$. The black hole mass estimates of \iras\, and \twomas\, are log\,M$\rm_{BH}$\,=\,8.59\,$\pm$\,0.14 (M$_\odot$) and log\,M$\rm_{BH}$\,=\,8.21$\pm$0.2 (M$_\odot$), respectively. In the X-ray bands, \iras\, is compatible with a Type 1 object, showing both spectral and flux variability. \chandra\, data of 2MASX\,J06021107+2828382 allowed us to measure a high hardness ratio in this source, providing evidence for a Type 2 AGN. The 22 GHz image obtained with the {\it Karl G. Jansky Very Large Array} has revealed that both AGN are compact radio objects with spectral indices -0.26$\pm$0.03 and -0.70$\pm$0.11, confirming for the first time its dual AGN nature in the radio bands. 
\end{abstract}


 

\section{Introduction}

In the cosmological $\Lambda$CDM paradigm, galaxies are proposed to grow hierarchically via mergers \citep{1978MNRAS.183..341W,1991ApJ...379...52W}. Galaxy mergers are mechanisms that can produce major gas inflows towards the galactic centres that can trigger bursts of star formation and also be responsible for the supermassive black hole (SMBH, with masses in the range of 10$^{6}$ to 10$^{10}$\,M$\odot$) growth in AGN  
\citep[][]{1991ApJ...370L..65B,2005Natur.433..604D,2008ApJS..175..356H,2018MNRAS.479.3952B}. 

Observational studies have led to the discovery that nearly every massive galaxy with a bulge harbours a SMBH in its centre. SMBHs have fundamental connections with the evolution of the host galaxy and the SMBH growth \citep[][]{1998AJ....115.2285M,Gebhardt_2000,Ferrarese_2000,2002ApJ...574..740T,2005SSRv..116..523F,2013ARA&A..51..511K,2009ApJ...698..198G,
2013ApJ...764..184M,2019ApJ...887..245S}, as well as the evolution of galaxies during mergers. Therefore, if SMBHs are ubiquitous in galaxy centres, a natural consequence is to expect that merging and merged galaxies will host two, or more, SMBHs \citep[][]{1980Natur.287..307B,2001ApJ...563...34M}. Thus, a large fraction of them should be found in many sources  \citep[][]{2012ApJ...748L...7V, 2019ApJ...887...90L,2020ApJ...904...23K}.

The so-called dual AGN (DAGN) are galaxies in a late merger phase that lasts $\sim$100\,Myr  where close interactions produce that two AGN are powered by accretion onto a SMBH with projected separations in the range of a few $\sim$kpc \citep{2009ApJ...698..956C,2013MNRAS.429.2594B}, although some authors accordingly with simulations \citep[e.g.,][]{2012ApJ...748L...7V} used to define DAGN as systems with their stellar bulges separated by $<$10 kpc \citep[e.g.][]{2021ApJ...923...36S}. 
Since gas-rich mergers could be obscured by dust, \citep{2018ARA&A..56..625H} DAGN are difficult to detect in the optical bands and are considered as DAGN candidates until the pair of central SMBHs are confirmed with either X-ray or radio observations \citep[][]{2015ApJ...815L...6F}. Since more than a decade, several studies dedicated to the detection of DAGN have been done in the optical, \citep[e.g.,][]{2010ApJ...715L..30L,2014MNRAS.437...32W,2020ApJ...904...23K}; mid-infrared (MIR) 
\citep[e.g.,][]{2017ApJ...848..126S,2017MNRAS.470L..49E}; radio \citep[e.g.,][]{2015ApJ...799...72F,2015ApJ...813..103M,2019MNRAS.484.4933R}; and in X-ray
\citep[e.g.,][]{2003ApJ...582L..15K,2019ApJ...887...90L,2019ApJ...882...41H}. In this context, binary-AGN are defined as a pair of active SMBHs which are forming a Keplerian binary bound system with pc to sub-pc projected separation  \citep[see review by][]{2019NewAR..8601525D}.

The detection and characterization of dual and binary SMBHs is fundamental to understand the formation and accretion history of SMBHs across cosmic ages. Until recently, a few dozens of DAGN were confirmed as such in the literature \citep[e.g.][]{2018JApA...39....8R}. However, the number of DAGN is increasing. For example, \citet{2021AJ....162..289Z} report 16 likely DAGN and 15 new identified DAGN in a systematic search for DAGN among 41 merging galaxies separated at $\sim$kpc scales at z$\leq$0.25. Also, \citet{2022MNRAS.512L..27B} find a new sub-kpc binary-AGN in the radio bands with a projected separation of 0.95$\pm$0.29 kpc. 

In particular, X-ray bands above 2 keV are found to be extremely useful to identify DAGN, since AGN with multiple nuclei are strong hard X-ray emitters, which also are less affected by the emission produced by the host galaxy \citep[e.g.][]{2003ApJ...582L..15K,2011ApJ...735L..42K}. Among the DAGN detected in hard X-ray, the serendipitous discovery by \citet[][]{2003ApJ...582L..15K} of the ongoing merger of two disk galaxies in NGC\,6240 using \textit{Chandra} observations stands out. This object, is known to have double AGN separated by $\sim$1\arcsec 
(or $\sim$0.7\,kpc) in projection. Observations in the radio and near-IR bands by \citet{2007Sci...316.1877M} also confirmed its dual nature. Recently, 3D spectroscopy observations revealed a third nucleus in a region of only 1\,kpc in this source, i.e., a triple-AGN candidate system \citep[][]{2020A&A...633A..79K}. However, revisiting the \textit{Chandra} data, it was discovered that the third nucleus in NGC\,6240 does not have evidence of hard X-ray emission \citep[][]{2020ApJ...902...49F}.

Initially, \citet{2004ApJ...604L..33Z} suggested that a method for finding DAGN could be based on the detection of double-peaked narrow emission line profiles (DPAGN) in the AGN spectra. Since then, hundreds of DPAGN were detected using the Sloan Digital Sky Survey database \citep[SDSS; e.g.][]{2012ApJ...745...67F,2012ApJ...753...42C,2020ApJ...904...23K}.
However, DPAGN are sometimes  confirmed as DAGN systems, but frequently can be the result of gas kinematics, jet-driven outflows, winds or disk rotation on small scales 
\citep[e.g.,][]{2011ApJ...727...71F,
2000ApJ...532L.101C,
2010ApJ...716..131R,2017ApJ...846...12K,2015ApJ...813..103M,2018ApJ...867...66C,2018MNRAS.474L..56B,2019MNRAS.490.5521B}. 

\iras\, is a nearby AGN at z=0.033  \footnote{The scale with 
our adopted cosmology is 0.662 kpc/", see Table \ref{table:obs}.} 
that was detected with the \textit{Swift}-Burst Alert telescope \citep[BAT;][]{Barthelmy2005} and catalogued as a Type 1 hard X-ray source. In the optical bands, a spectroscopy study done by \citet[][]{2010ApJ...710..503W} classify \iras\, as a Seyfert 1 galaxy and estimated a SMBH mass of  log\,M$_{BH}$=8.22$^{+0.01}_{-0.01}$ (M$_\odot$). Moreover, published data and the spectra presented in this paper allows locating \iras\, along the quasar main sequence \citep[e.g., ][]{sulenticetal00a,shenho14}, a correlation space that helps the contextualization of Type 1 AGN observational properties, and that provides a tentative link to physical and aspect parameters \citep{sulenticetal11}. 

In a previous study done by \citet{2010ApJ...716L.125K} 
they suggested that the companion galaxy of \iras\, is 2MASXJ\,06021038+2828112, having a projected separation of 8\,kpc. Later, \citet{2012ApJ...746L..22K} found that \iras\, is a DAGN based on the analysis of X-ray data obtained with \textit{Chandra}. Specifically, in their Table 1, \iras\, appears as a DAGN system conformed by a pair of Seyfert 2 (Sy2) galaxies, being the companion galaxy \twomas, and both with a projected separation $\sim$ 8 kpc. Since then, \iras\, is known as a  confirmed DAGN in the X-ray by direct imaging with a projected separation less than 10\,kpc \citep[see, e.g.][]{2014Natur.511...57D,2017ApJ...848..126S}.

In order to provide a detailed study of the nature of the DAGN system previously detected in \iras,  new optical spectroscopic data were obtained along with data from radio and X-ray databases. Our data analysis shows that the galaxy pair conformed by \iras\, and \twomas, is a DAGN in the radio bands. In Section~\ref{optical} the optical spectra obtained with three telescopes are presented; while section~\ref{model} shows the analysis done to the optical data. In Section~\ref{radio} analysis of the \vla\,data of \iras\, and \twomas\, is given. In section~\ref{X-ray} X-ray data for both AGN are presented and analysed. Finally, a discussion of the obtained results in the optical, radio and X-ray is given in Section~\ref{dis}. Throughout the paper we adopt a cosmology where  $\rm{H_{0}}$=69.6~km\,s$\rm{^{-1}}$\,Mpc$\rm{^{-1}}$, $\rm{\Omega_{m}}$=0.286 and $\rm{\Omega_{\lambda}=}$0.714 \citep{2014ApJ...794..135B}.

\section{Optical spectroscopic observations}
\label{optical}

A set of new optical spectroscopic observations were obtained with the aim to study the dual AGN \iras\, system previously reported as such by \citet{2012ApJ...746L..22K} and also the spectral properties of some of its close neighbours. 

\subsection{\textit{WHT/ISIS}--(ORM) }
\label{WHT}

Optical long-slit data of \iras\, were obtained using the Intermediate dispersion Spectrograph and Imaging System (ISIS), attached to the 4.2\,m William Herschel Telescope (\textit{WHT}) at the Roque de los Muchachos Observatory (ORM). ISIS has two CCD arrays, an EEV12 for the blue arm and a  REDPLUS CCD for the red arm. The blue CCD was centred around 4500\,\AA, and the red one at 6999\,\AA. The gratings used were R600B and R600R, which provide a dispersion of 0.44 and 0.49\,\AA\,pixel$^{-1}$, respectively. The slit width was set to 1.018\,\arcsec, this is about 3.3 pixels FWHM. The spectral resolution in the blue region is 2.06\,\AA\, and in the red 1.84\,\AA. The slit was placed at a position angle (PA) $\sim$-92$^{\circ}$ and observations were done with an average seeing 1.2\arcsec. With this setup, three exposures of 1200\,s in the blue (3550-5250\,\AA), and in the red (5860-7780\,\AA) were obtained. Data were reduced and calibrated using standard \textsc{IRAF} packages\footnote{\textsc{IRAF} is distributed  by the  National Optical Astronomy Observatories, operated by the Association of Universities for Research in Astronomy, Inc., under cooperative  agreement with  the  National Science Foundation.}. For the wavelength calibration, CuAr and HeNeAr lamps were used. The standard star Feige\,34 from the \citet{1990AJ.....99.1621O} catalogue was used for flux calibration. Sky subtraction was done using task \textsc{IRAF/Background}. Spectra were combined to produce a final 3600\,s exposure time. An aperture of 3\arcsec\, was used to extract the 1-dimensional spectrum. The obtained \textit{WHT} spectrum has a signal-to-noise ratio (S/N)\,$\sim$55 in the continuum around, 5100 \AA. The complete log of optical observations obtained with \textit{WHT/ISIS}, \textit{2.1m/B\&Ch} and \textit{Copernico/AFOSC} is presented in Table~\ref{table:obs}.  

\subsection{ \textit{2.1m/B\&Ch}--(OAN-SPM) }
\label{OAN-SPM}

Optical spectroscopic observations of \iras\, and some of its close neighbours were obtained on 2018 October 18,with the 2.1 m telescope and the Boller \& Chivens spectrograph (\textit{2.1m/B\&Ch}) at the Observatorio Astron\'omico Nacional at San Pedro M\'artir, Baja California, Mexico (OAN-SPM). The \textit{2.1m/B\&Ch} was used with a 300 l/mm grating, covering the spectral range of 3800-8000\,\AA. The slit width was set to 2.5\arcsec\, and the PA was oriented E-W. The spectral dispersion is 4.5\,\AA\,pix\,$^{-1}$, corresponding to 10\,\AA\, FWHM, which was estimated using different emission lines of the arc-lamp (CuHeNeAr) spectrum. Observations were obtained with clear sky conditions and a seeing of 2\arcsec.  Data were reduced with \textsc{IRAF} following standard procedures. Arc-lamp (CuHeNeAr) exposures were used for the wavelength calibration. A spline function was fit to determine the dispersion function (wavelength-to-pixel correspondence). Flux calibrations were done using the spectrophotometric standard star Feige\,110 from  \citet{1990AJ.....99.1621O} which was observed each night.

The spectra obtained for \iras\, and some of its neighbours along with the identification map from the Digitized Sky Survey (DSS, \footnote{\url{https://archive.eso.org/dss/dss}}) of all observed sources are shown in the left and right panels of Figure~\ref{Fig:all-SPM}, respectively. Spectrum in panel ({\it a}) corresponds to \iras; panel ({\it b}) shows the spectrum of the extended emission located at the NE of \iras; in ({\it c}) 2MASX\,J06021107+2828112 shows a stellar-like spectrum with some faint emission lines that probably are due to extended emission at the SW of \iras; panel ({\it d}) shows the spectrum of 2MASSX J06021107+2828382 where AGN emission lines are evident; finally, panel ({\it e}) shows the stellar-like spectrum of 2MASX\,J06021107+2828269. 

\subsection{{\textit Copernico/AFOSC}--(INAF-OAPd)}
\label{Asiago}

Observations of \iras\, were obtained with the 182 cm {\it Copernico} telescope at the Astronomical Observatory of Padova (INAF OAPd - Asiago Observatory) and the Asiago Faint Object Spectrograph and Camera (\textit{AFOSC}). The aim for this new set of optical spectroscopic observations was to study the low brightness extended emission initially observed in the \textit{PanSTARSS} image of\, \iras
\footnote{image available at http://cdsportal.u-strasbg.fr}, see upper panel of Figure~\ref{Fig:Pans-halfa}, between \iras\, and \twomas. Henceforth, we will refer to this extended emission as the ``bridge'' zone. Data were obtained with the GR08 Grism with a central wavelength $\lambda_{cen}$=7000\,\AA, and a spectral range from 6250 to 8050\,\AA. Neon and Thorium arc-lamps were used and combined for wavelength calibration. The spectral dispersion is 1.98\,\AA\,pix\,$^{-1}$, corresponding to 3.3\,\AA\, FWHM, which was estimated using the arc-lamp spectrum. The slit aperture was set to 1.69\arcsec. The final spectra were calibrated in flux using the spectrophotometric standard star PG 0205+134 also from \citet{1988ApJ...328..315M}. Two images of 3600\,s were obtained with a PA=21$^{\circ}$.

\subsection{The bridge zone}
\label{bridge}

The 2D \textit{Copernico} spectrum shows evidence of an extended H$\alpha$ emission in the bridge zone, see lower panel of  Figure~\ref{Fig:Pans-halfa}. So, using this 2D spectrum, an estimation of the  size of the extended emission was done. This resulted in a region of $\sim$15 pixels. Since the scale for this setup is 0.5\arcsec\,pix$^{-1}$ for a binning of 2$\times$2, this corresponds to $\sim$4.965 kpc. 

In order to verify this result, deep spectra of the bridge zone were obtained also with the \textit{2.1m/B\&Ch} at the (OAN-SPM) on 2020 February 26. Observations were obtained  with clear sky conditions and a seeing of 2.5\arcsec. The slit has a PA$\sim$21$^{\circ}$ in order to cover the bridge zone and again with an aperture of 2.5\arcsec. Spectra were processed using the same procedure explained in section~\ref{OAN-SPM}. The spectra were extracted using an aperture of 4 pixels (i.e., 2.4\arcsec). The flux calibration was done using the spectrophotometric standard star Feige\,34 from the \citet{1990AJ.....99.1621O} catalogue. Three spectra of 1800\,s of integration time were combined to obtain a final 1.5 hr 2D spectrum. Evidence of extended emission from [OIII]$\lambda\lambda$4950,5007 in \iras\, is also found, see Figure~\ref{Fig:extended}. 


\begin{table*}
\centering
\caption{\bf Log of optical spectroscopy observations}
\label{table:obs}
\begin{tabular}{lcccccc}
\hline 
\hline \\
Telescope/Instrument & Object ID & Date (UT) & Exposure time (s) &  z$^{a}$ & S/N$^{b}$\\
\hline \\
 \textit{WHT/ISIS} & \iras\, & 2015 January 28  & 3600 & 0.0330 $\pm$ 0.0002 & 53 \\ 
\hline \\
\textit{2.1m/B\&Ch} & \iras\, & 2018 October 18  & 900 & 0.0329 $\pm$ 0.0001 & 54 \\
& \iras\, NE            & '' & 900  & 0.0334 $\pm$ 0.0001 & 6\\
& 2MASXJ06021107+2828382 & '' & 5400 & 0.0334 $\pm$ 0.0001 & 28\\
& 2MASXJ06021038+2828112       & '' & 3600$^{c}$ & 0.0320 $\pm$ 0001 & 40 \\
& 2MASSJ06021215+2828269       & '' & 1800$^{c}$ & {star}& 61\\
& The bridge & 2020 February 26 & 5400 \\
\hline \\
\textit{Copernico/AFOSC}\\
& The bridge & 2019 January 10 & 3600 \\
\hline \\
\multicolumn{4}{l}{$^{a}$ Redshifts of \iras\, and its close neighbors were estimated using narrow emission lines.}\\
\multicolumn{2}{l}{$^{b}$ Measured around 5100\AA}\\ 
\multicolumn{2}{l}{$^{c}$In the same slit.}\\
\end{tabular}
\end{table*}

\begin{figure*}
\begin{center}
{\includegraphics[width=17.5cm]{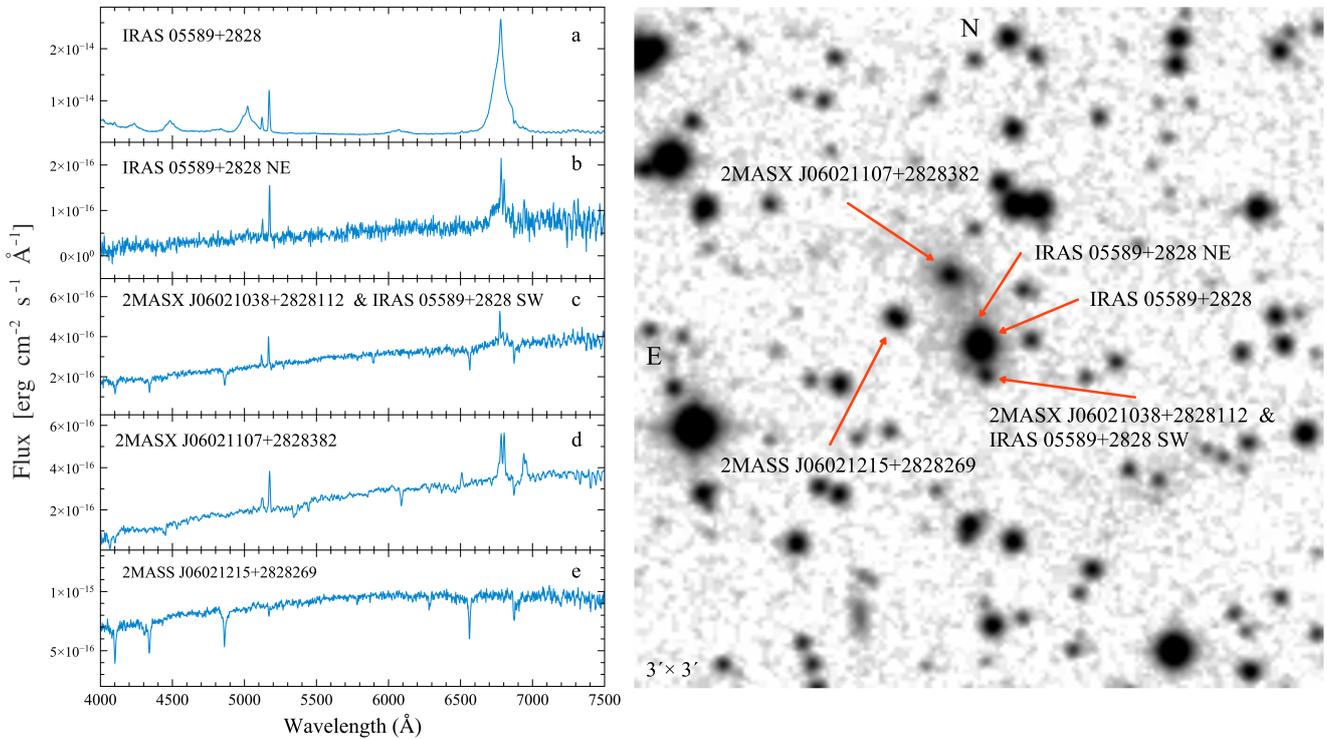}}
\caption{Left: Optical calibrated spectra of \iras\, and its close neighbours were obtained with the \textit{2.1m/B\&Ch}. Right: DSS (plate O) used as an identification map for \iras\, and its close neighbours. Red arrows mark the positions and names of the observed sources.} 
\label{Fig:all-SPM}
\end{center}
\end{figure*}


\begin{figure}
\begin{center}
{\includegraphics[width=8.9cm]{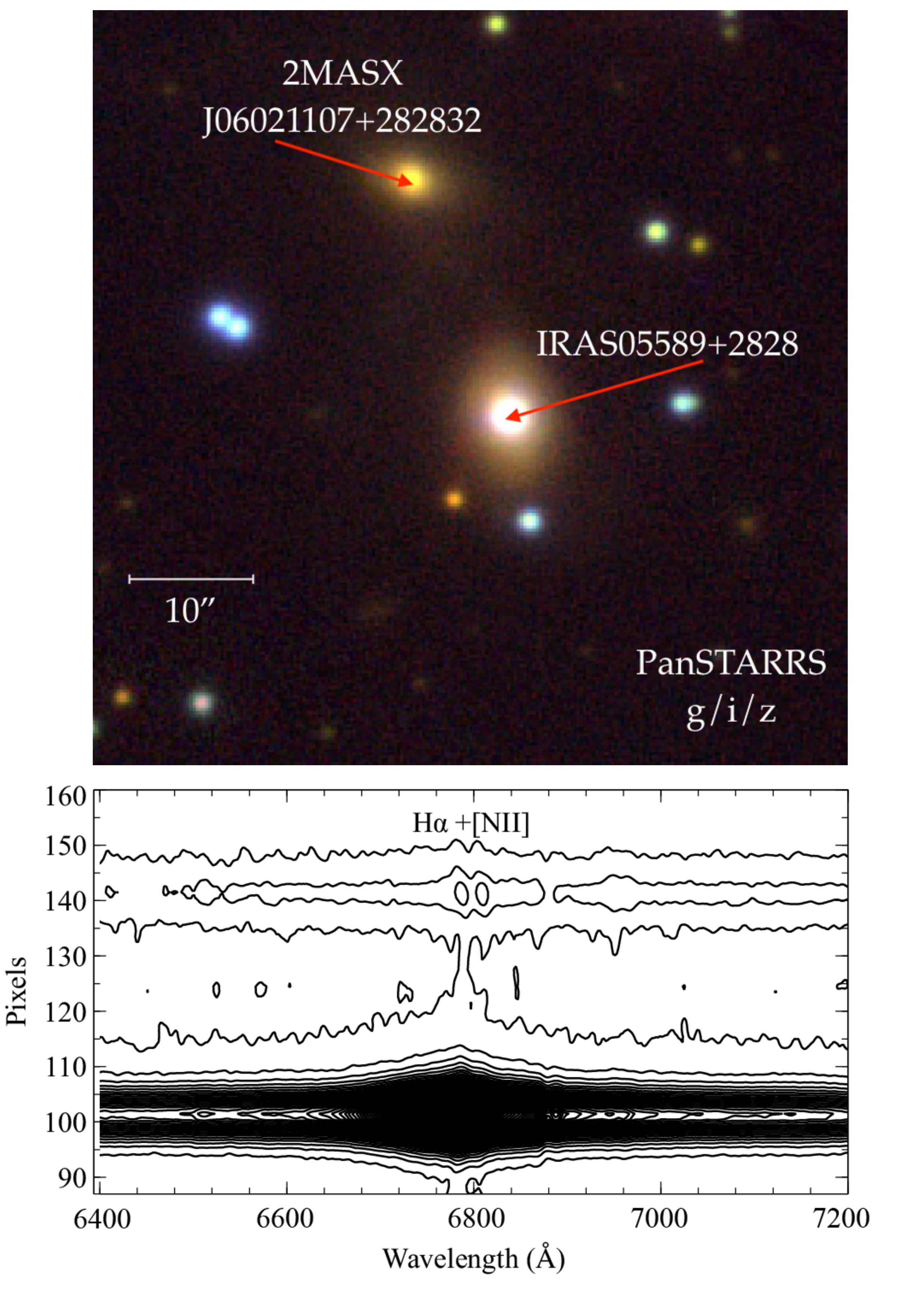}}
\caption {Upper panel: \textit{PanSTARRS}/DR1 composite {\it g}, {\it i} and {\it z} bands image of \iras\, and \twomas. 
Amidst the two objects, hints of low surface gas can be observed in the bridge zone. Lower panel: A 2D contour map of the flux calibrated spectra of bridge zone was obtained with the \textit{Copernico} telescope on 2019 January 10. Extended emission from H$\alpha$ is shown.}
\label{Fig:Pans-halfa}
\end{center}
\end{figure}

\begin{figure}
\begin{center}
{\includegraphics[width=7.75cm]{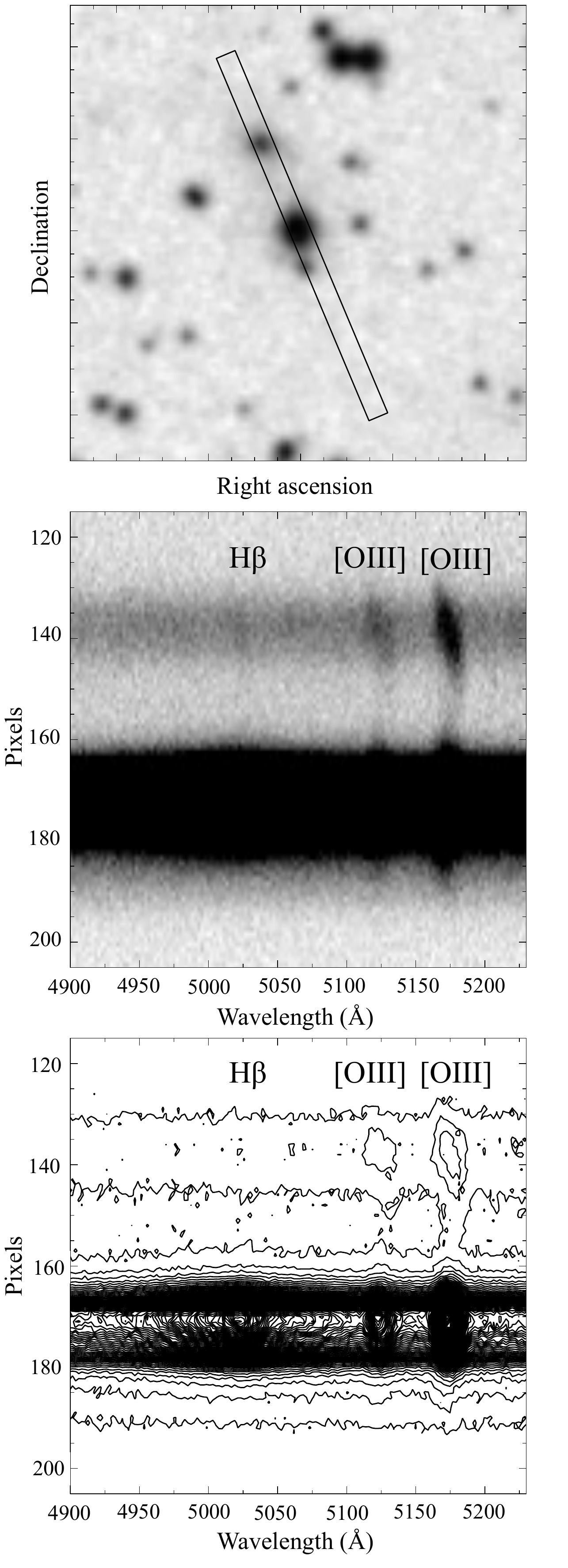} }
\caption{Upper panel shows the slit position used to study the bridge zone with the \textit{2.1m/B\&Ch} at OAN-SPM. Middle panel shows the [OIII]$\lambda\lambda$\,4959,5007 \AA\, extended emission between \iras\, and \twomas, detected in the 2D spectrum. The bottom panel shows a contour map of the bridge zone.}
\label{Fig:extended}
\end{center}
\end{figure}
\section{Emission line profiles models}
\label{model}

\subsection{IRAS05589+2828}
\label{model1}

\begin{figure*}
\begin{center}
{\includegraphics[width=14cm]{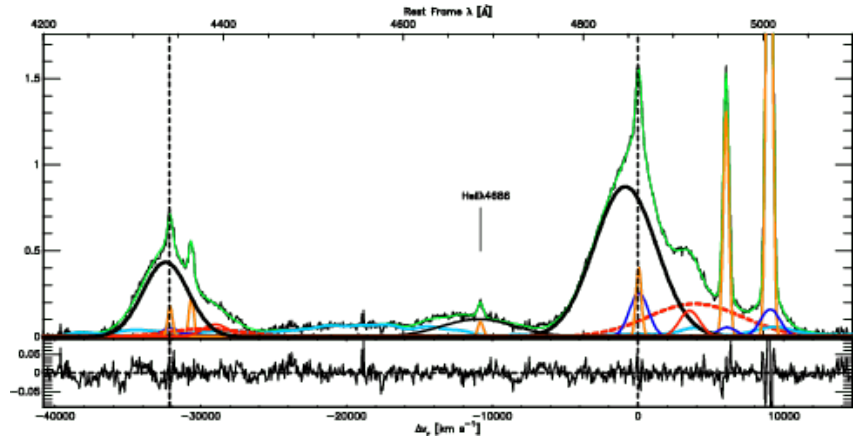}}
\caption{The thin black line shows the original \textit{WHT/ISIS} spectrum of \iras, obtained after subtracting the continuum emission. The emission line profiles are shown around two main spectral regions. Vertical dashed lines mark the positions of the rest frames of \hg\, and H$\beta$. The complex broad Balmer emission lines profiles were fitted with three Gaussian components. The thick black line shows the blueshifted Gaussian main component (MC); the redshifted very broad component (VBC) is fitted with a Gaussian component shown with a dashed red line; a third Gaussian component used to fit the red-peaklet (RP) is shown with a red continuous line. The narrow lines were best fitted using two Gaussian components, shown in orange and dark blue. Both components are located in the restframe. The FeII pseudo-continuum emission was model using a template that is shown with a cyan line. The emission of HeII$\lambda$4686 was also included in the model, but with a single broad component due to its low $S/N$ ratio. The best fit obtained is shown in green. The same colours are used for H$\gamma$ and [OIII]$\lambda$4363 spectral regions.The bottom panel shows the obtained residuals.} 
\label{Fig:all-hbeta}
\end{center}
\end{figure*}

\begin{figure}
\begin{center}
{\includegraphics[width=7cm]{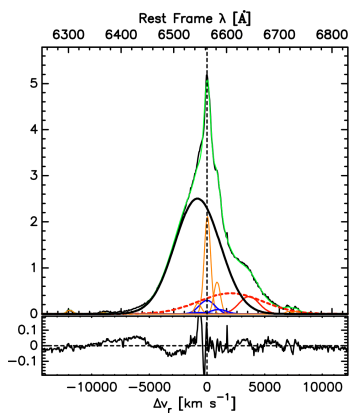}}
\caption{Best fit obtained around the H$\alpha$ region for \iras. The broad emission lines profiles were model using three Gaussian components, following the procedure done in Figure~\ref{Fig:all-hbeta}. There is no FeII emission contribution in the red spectral region. The bottom panel shows the residuals.} 
\label{Fig:all-halfa}
\end{center}
\end{figure}

The AGN spectrum of \iras\, obtained with \textit{WHT/ISIS} was modelled with \textsc{IRAF/Specfit}. This routine allows fitting the emission lines and the underlying continuum at the same time \citep[][]{1994ASPC...61..437K}. The spectrum was modelled in the blue and red spectral regions.
The blue region (4230-5070\,\AA) includes \hg\ and \hb\ (see Figure~\ref{Fig:all-hbeta}).
The red region (6250-6850\,\AA) includes only \ha+\nii\ (see Figure~\ref{Fig:all-halfa}).   
For each spectral region we considered continuum windows at 4250, 4450 and 5050 \AA\ for the \hb+\hg\ region and at 6150 and 6850 \AA\ for H$\alpha$+N[II]. The first step consisted in modelling the underlying continuum with a power law, taking as a reference the windows described above. Then, the fitting was done leaving as free parameters the flux, velocity shift and FWHM. 

\begin{figure}
\begin{center}
\subfigure{\includegraphics[width=6cm]{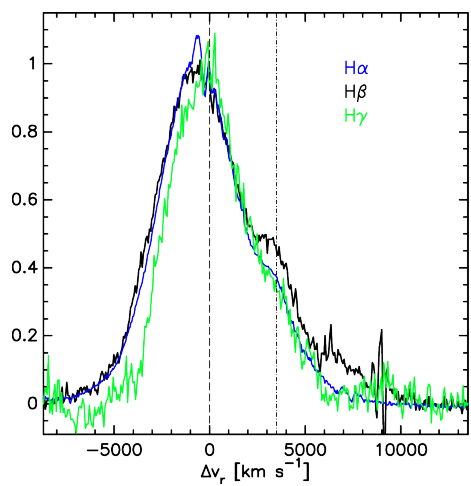}}
\caption{Profile comparison of the main broad Balmer emission lines obtained with normalized fluxes for \iras. The observed spectra of \ha, \hb, and \hg\, are superimposed after subtracting the FeII and the narrow components. The dashed vertical line shows the systemic restframe wavelength. A dot-dashed vertical line marks the position of the RP that has a velocity shift of ($\Delta V_{r}$)\,$\sim$\,3500 \kms.}
\label{Fig:profile}
\end{center}
\end{figure}

The best fit for the Balmer broad line profiles was obtained using three broad Gaussian components. The first Gaussian was used to fit the most intense component, identified as the main component (MC), which is blueshifted with respect to the restframe. This component is interpreted as emerging from the broad line region, the closest to the SMBH. The width of the line is assumed to be due to Doppler broadening of the clouds in a co-planar geometry \citep[e.g.][]{Netzer2013}. 
The second Gaussian component was used to fit a redshifted very broad component (VBC), a component frequently observed in objects with FWHM > 4000 \kms\ and low FeII emission  \citep[c.f.,][]{marzianietal10,zamfiretal10}. Finally, a third Gaussian component was used to fit a redshifted peaked emission that has a velocity shift of $\sim$3500 \kms\ with respect to the restframe, see Figure~\ref{Fig:profile}. Henceforth, we will refer to this peculiar component as the red-peaklet (RP). The results of the fitting of these components MC, VBC, and RP are shown in Table~\ref{tab:Tab_fit}. A FeII template was used to model the pseudo-continuum. Also, modelling the HeII$\lambda$4686 emission line was necessary in order to obtain the best fit. 

The best fit for the narrow emission lines was obtained using two Gaussian components. This fit was done using the same FWHM and velocity shift for all lines, based on the assumption that these lines are produced in the same region. The fitted narrow lines are: \ha, \nii, \sii, \oi, \oiiill, \hb, \heii, \hg, and \oiiig. Both narrow components are centred in the restframe, with an average shift of $\pm$1\,\AA. 

Therefore, in this object there are no blueshifted nor redshifted components as it is usually the case in DPAGN. So, this object has a double narrow emission line component \citep[e.g.,][]{2006MNRAS.371.1610I}. The first narrow component (NC), which is the most prominent, has a FWHM $\sim$\,500\,\kms. For the second component (SC), similar to what was done with the broad emission lines, a Gaussian component with FWHM $\sim$ 1550 \kms, was used. 
The second component is expected in quasars, particularly in \oiiill\ which is a high-ionization line that is frequently affected by broadening and shifts with respect to the quasar rest frame due to disk winds \citep[e.g.,][]{zamfiretal10,Zhang2011}.
However, we assume that a similar emission could emerge from the other narrow lines if they share the same emitting region. The SC is well visible in the strongest lines like \ha, \hb, and \oiiill, and appears weaker in \nii, \hg, and \oiiig, while no emission from the SC is observed in the faintest lines like \heii, \sii, and \oi.

Results of the best fit for the broad and narrow emission lines profiles of \iras\, are presented in Table~\ref{tab:Tab_fit}. The presence of broad permitted lines in the optical spectrum of \iras\, confirms that it is a Sy1 AGN, in agreement with previous results \citep[e.g.][]{2013ApJS..207...19B}. 

\begin{table*}
\caption{\bf Profile modelling results of \iras}
\label{table:irasmodel}
\begin{tabular}{lcccccc}
\hline \hline \\
\textit{WHT} & Flux\,$\times$10$^{-14}$ & Luminosity\,$\times$10$^{41}$ & $\lambda_{cent}^a$ & $\Delta$v$^b$  & FWHM \\
             & erg\,s$^{-1}$\,cm$^{-2}$\,\AA$^{-1}$ & erg\,s$^{-1}$ & \AA &\kms &\kms \\
\hline \\
\hg $_{\rm MC}$	&	27.47	$\pm$	0.97	&	6.87	$\pm$	0.12	&	4335.7	$\pm$	0.5	&	-294	$\pm$	33	&	4115	$\pm$	86	\\
\hg $_{\rm RP}^{c}$	&	4.94	$\pm$	1.30	&	1.24	$\pm$	0.16	&	4380.6	$\pm$	7.8	&	2803	$\pm$	536	&	6713	$\pm$	824	\\
\hgnc	&	1.22	$\pm$	0.08	&	0.30	$\pm$	0.01	&	4340.7	$\pm$	0.1	&	46	$\pm$	7	&	446	$\pm$	3	\\
\hgsb	&	1.22	$\pm$	0.08	&	0.30	$\pm$	0.01	&	4340.8	$\pm$	0.3	&	58	$\pm$	20	&	1540	$\pm$	102	\\
\hgvbc	&	2.61	$\pm$	0.54	&	0.65	$\pm$	0.07	&	4391.0	$\pm$	2.1	&	3521	$\pm$	142	&	2330	$\pm$	302	\\
\oiiig	&	1.54	$\pm$	0.09	&	0.39	$\pm$	0.01	&	4364.3	$\pm$	0.2	&	91	$\pm$	14	&	446	$\pm$	3	\\
\oiiig$_{\rm SC}$	&	0.55	$\pm$	0.03	&	0.14	$\pm$	0.00	&	4363.8	$\pm$	0.3	&	58	$\pm$	20	&	1540	$\pm$	102	\\
\hline 
\hb	$_{\rm MC}$ &	75.72	$\pm$	3.21	&	18.95	$\pm$	0.40	&	4846.8	$\pm$	0.3	&	-875	$\pm$	20	&	5043	$\pm$	27	\\
\hb $_{\rm RP}$	&	4.34	$\pm$	0.35	&	1.09	$\pm$	0.04	&	4918.2	$\pm$	0.7	&	3530	$\pm$	42	&	1636	$\pm$	92	\\
\hbnc	&	3.09	$\pm$	0.17	&	0.77	$\pm$	0.02	&	4861.7	$\pm$	0.1	&	46	$\pm$	7	&	446	$\pm$	3	\\
\hbsb	&	6.71	$\pm$	0.59	&	1.68	$\pm$	0.07	&	4861.9	$\pm$	0.3	&	58	$\pm$	20	&	1540	$\pm$	102	\\
\hbvbc	&	25.19	$\pm$	4.19	&	6.30	$\pm$	0.52	&	4924.4	$\pm$	7.2	&	3908	$\pm$	443	&	7479	$\pm$	649	\\
\oiiib	&	30.79	$\pm$	0.25	&	7.71	$\pm$	0.03	&	5006.9	$\pm$	0.0	&	-8	$\pm$	1	&	446	$\pm$	3	\\
\oiii$_{\rm SC}$	&	4.39	$\pm$	0.36	&	1.10	$\pm$	0.04	&	5008.0	$\pm$	0.3	&	58	$\pm$	20	&	1540	$\pm$	102	\\
\hebc	&	10.52	$\pm$	0.61	&	2.63	$\pm$	0.08	&	4683.8	$\pm$	1.7	&	-144	$\pm$	111	&	6178	$\pm$	292	\\
\henc	&	0.65	$\pm$	0.07	&	0.16	$\pm$	0.01	&	4685.8	$\pm$	0.4	&	-10	$\pm$	25	&	446	$\pm$	3	\\
\hline 
\ha	$_{\rm MC}$ &	270.86	$\pm$	5.59	&	67.78	$\pm$	0.70	&	6544.8	$\pm$	0.1	&	-830	$\pm$	7	&	4657	$\pm$	28	\\
\ha$_{\rm RP}$	&	18.09	$\pm$	1.28	&	4.53	$\pm$	0.16	&	6642.8	$\pm$	0.6	&	3643	$\pm$	29	&	2033	$\pm$	15	\\
\hanc	&	32.58	$\pm$	0.44	&	8.15	$\pm$	0.05	&	6563.7	$\pm$	0.1	&	31	$\pm$	5	&	661	$\pm$	9	\\
\ha$_{\rm SC}$	&	11.08	$\pm$	0.15	&	2.77	$\pm$	0.02	&	6563.0	$\pm$	0.1	&	31	$\pm$	5	&	1619	$\pm$	162	\\
\havbc	&	68.51	$\pm$	5.70	&	17.14	$\pm$	0.71	&	6608.1	$\pm$	1.9	&	2061	$\pm$	86	&	6506	$\pm$	270	\\ \hline
\niib	&	10.70	$\pm$	0.38	&	2.68	$\pm$	0.05	&	6582.0	$\pm$	0.2	&	-138	$\pm$	11	&	661	$\pm$	9	\\
\niib$_{\rm SC}$	&	3.64	$\pm$	0.13	&	0.91	$\pm$	0.02	&	6585.0	$\pm$	0.0	&	-1	$\pm$	0	&	1619	$\pm$	162	\\
\siia	&	0.86	$\pm$	0.16	&	0.22	$\pm$	0.02	&	6718.0	$\pm$	0.1	&	31	$\pm$	5	&	661	$\pm$	9	\\
\siib	&	0.90	$\pm$	0.14	&	0.23	$\pm$	0.02	&	6732.0	$\pm$	0.1	&	31	$\pm$	5	&	661	$\pm$	9	\\
\oil	&	1.20	$\pm$	0.11	&	0.30	$\pm$	0.01	&	6302.6	$\pm$	0.7	&	126	$\pm$	35	&	661	$\pm$	9	\\
\hline \\
\multicolumn{5}{l}{$^{a}$ Central wavelength measured at the peak of the line.} \\
\multicolumn{5}{l}{$^{b}$ Velocity difference with respect to the systemic value. } \\
\multicolumn{2}{l}{$^{c}$ RP stands for red-peaklet. } \\
\end{tabular} 
\label{tab:Tab_fit}
\end{table*}

\subsection{The relativistic disk model}
\label{diskmodel}

A relativistic disk model was used to fit the broad emission line component of H$\rm{\beta}$ \citep[see][]{1989ApJ...339..742C,1989ApJ...344..115C}. The fit was meant to account for the RP in the profile observed at $\lambda\sim4918$\AA. This is the main peculiarity with respect to the typical Balmer line profiles of sources with FWHM H$\rm{\beta \gtrsim}$ 4000 \kms\,
\citep[e.g., ][Population B following \citealt{sulenticetal11}]{zamfiretal10}. 
In most cases, Population B objects show a prominent red wing or VBC  \citep[][]{sulenticetal02}, so it cannot be considered as a peculiarity.
The second characteristic of this class of AGN is associated with the core, whose centroid appears significantly blueshifted. 

\begin{figure*}
\begin{center}
{\includegraphics[width=8cm]{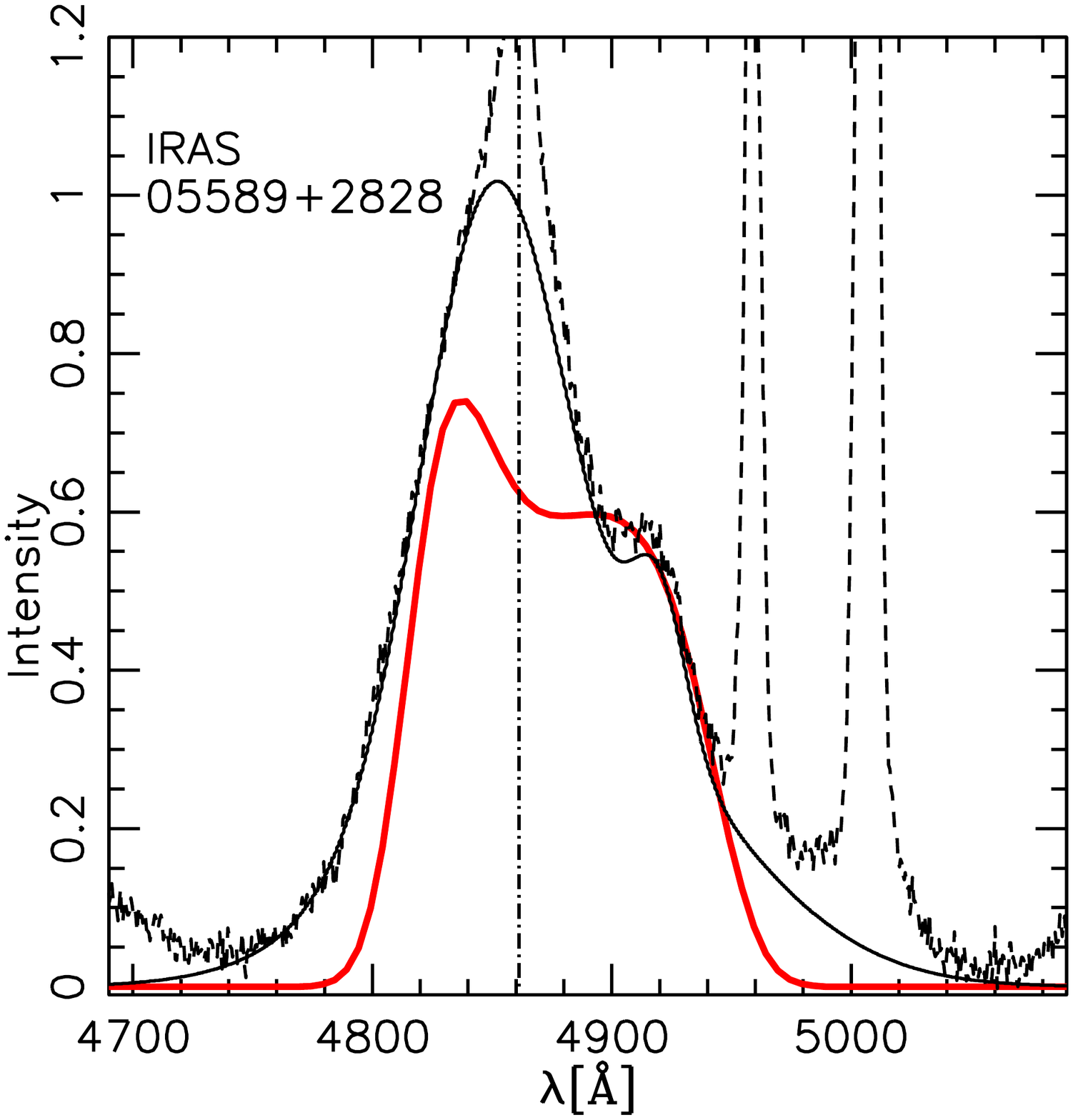}}
{\includegraphics[width=8cm]{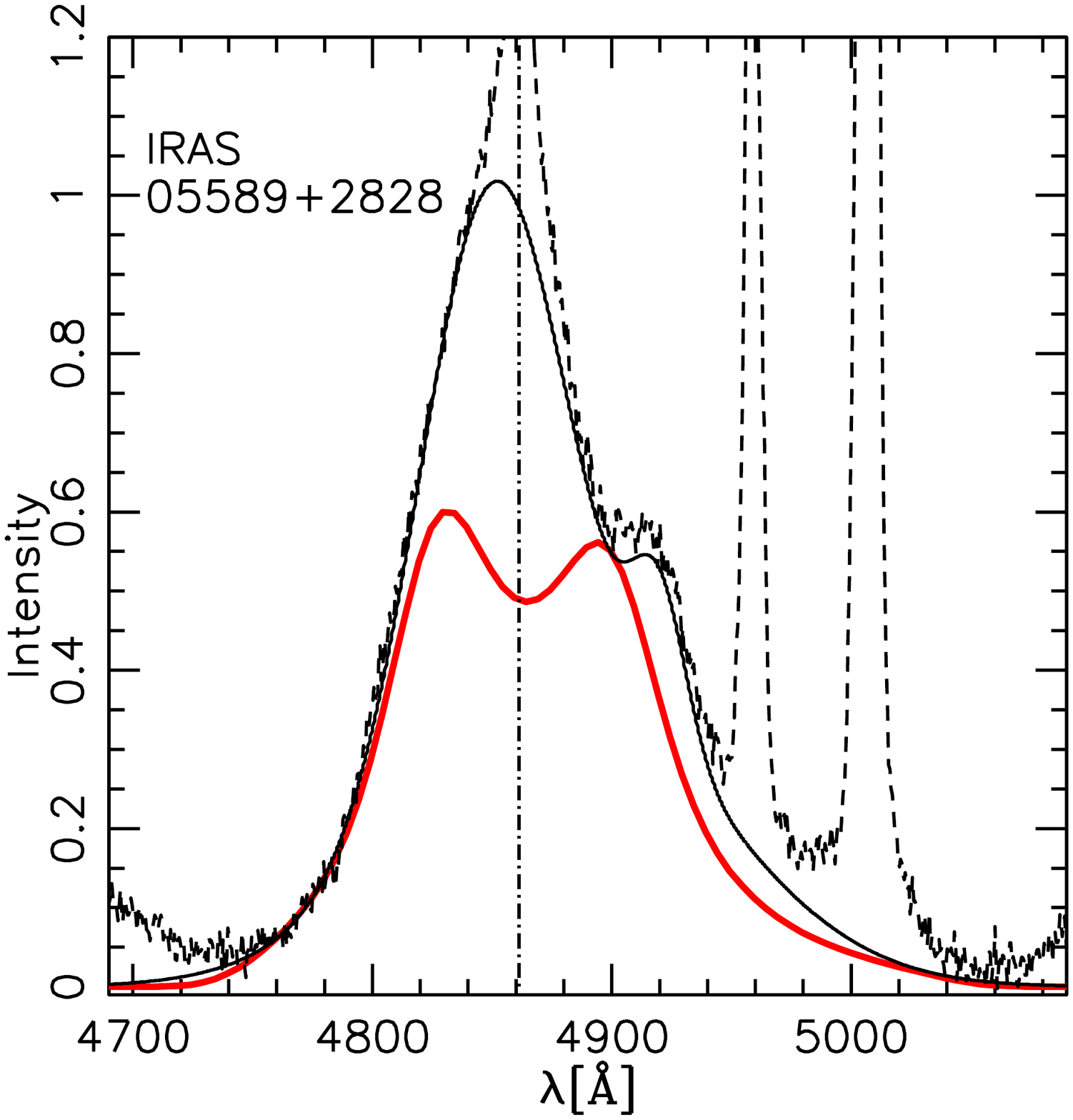}}
\caption{The broad H$\rm{\beta}$ profile (filled black line; the dashed line traces the continuum-subtracted spectrum)  modelled with a relativistic accretion disk (red line) of \iras. Left: disk model with parameters $\rm{\theta\,=\,15^{\circ}}$, $\rm{r_{in}\,=\,350}$\ $\rm{r_\mathrm{g}}$, $\rm{r_{out}\,=\,35000}$\ $\rm{r_\mathrm{g}}$, $q\,=\,3.6$, and turbulence parameter $\rm{\xi}$\,=\,2.8$\rm{\times10^{-3}}$. Right panel: disk model parameters are $\rm{\theta\,=\,28}$, $\rm{r_{in}\,=\,225}$\,$\rm{r_\mathrm{g}}$, $\rm{r_\mathrm{out}\,=\,4500}$\,$\rm{r_\mathrm{g}}$, $q\,=\,1.5$, and turbulence parameter $\rm{\xi\,=\,2.8\times\,10^{-3}}$.} 
\label{Fig:Disk}
\end{center}
\end{figure*}

The disk profiles shown in Figure~\ref{Fig:Disk} can account for part of the 
emission lines, as was found in previous works (see, for example \citealt{bonetal07,bonetal09a}). The left panel in Figure~\ref{Fig:Disk} shows that a disk profile seen at low inclination (viewing angle $\theta$\,=\,15$^{\circ}$, defined as the angle between the line-of-sight and the disk axis-of-symmetry) with a steep emissivity law as a function of the disk radius with $q = 3.6$ ($\rm{\epsilon \propto r^{-q}}$), and inner radius $\rm{r_{in} = 350}$\ $\rm{r_\mathrm{g}}$, can account for the RP. However, it clearly fails to reproduce the core as well as the extended red wing. Increasing the inclination and reducing the inner radius ($\rm{r_{in} = 225}$\ $\rm{r_\mathrm{g}}$) produces a fit that is able to reproduce fairly well the line base, nicely fitting the blue side and accounting for the shape of the red wing, but the RP is not reproduced. Therefore, the blue excess and the RP observed in the \hb\, profile of \iras\, should have a different origin. 

Since in all the optical spectra obtained with the three telescopes the RP is observed, in order to establish the possible origin of this feature, an estimation of a possible spatial separation between \iras\, and the RP is done using the 2D \textit{Copernico/AFOSC} red spectrum. A section of the 2D spectrum centred in the H$\alpha$ region of \iras\, was analysed  \citep[see][]{2011TJSAS..53..307S}. A series of Gaussian functions are fit line by line throughout the spectrum, estimating the FWHM in each one. Starting at the peak of the \ha\, emission, the collection of FWHM, with a mean of 5.18 pixels, were analysed. A small protuberance stands out in the middle of the noise at pixels 38-39, equivalent to a 75.15-77.13 \AA\, towards the red. This, in turn, is equivalent to 3435 - 3525 km\,s$^{-1}$, consistent with the adjustment of the RP mentioned before. The maximum value of this peak, representing a transversal widening of the spectra in the spatial direction, is only 0.1 pixel (0.096 $\pm$ 0.046) over the mean value of 5.18. This spatial size corresponds to 33 $\pm$ 16 pc. Certainly, this detection is marginal and purely statistical, being a small fraction of a pixel, where a full pixel covers 330 pc.


\subsection{2MASX J06021107+2828382}
\label{model2}

The optical spectrum of \twomas\, was obtained in 2018 October with the \textit{2.1m/B\&Ch} (see details in  Table~\ref{table:obs}). In order 
to model the stellar spectrum of the host galaxy of \twomas\, we use the \textsc{starlight} code  \citep[see][]{2005MNRAS.358..363C}.  The spectrum 
was corrected for Galactic extinction using the dust maps by 
\citet{1998ApJ...500..525S} and the extinction law of 
\citet{1989ApJ...345..245C} with $R_v = 3.1$. 

\begin{figure}
\begin{center}
{\includegraphics[width=8cm]{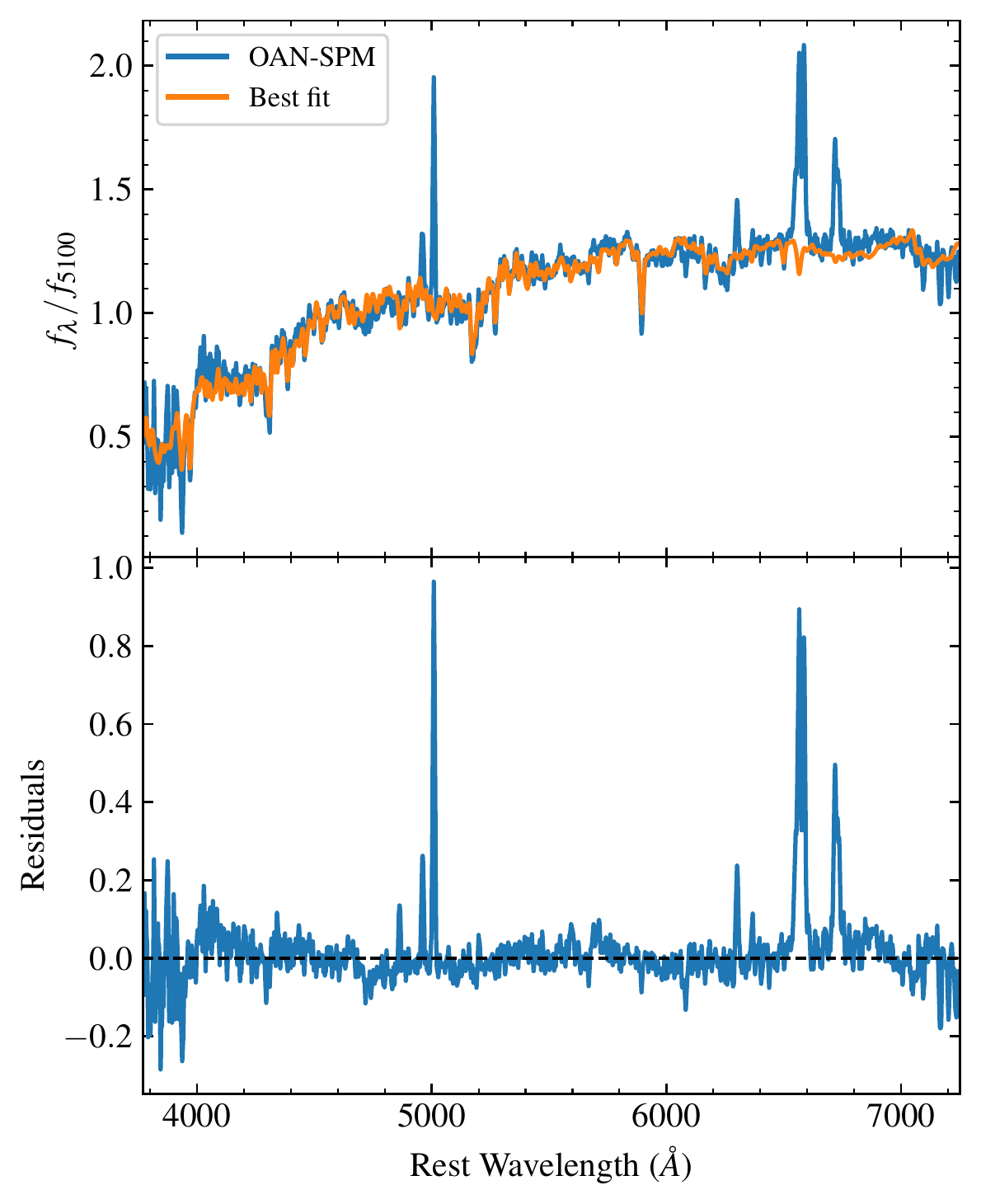}}
{\includegraphics[width=8cm]{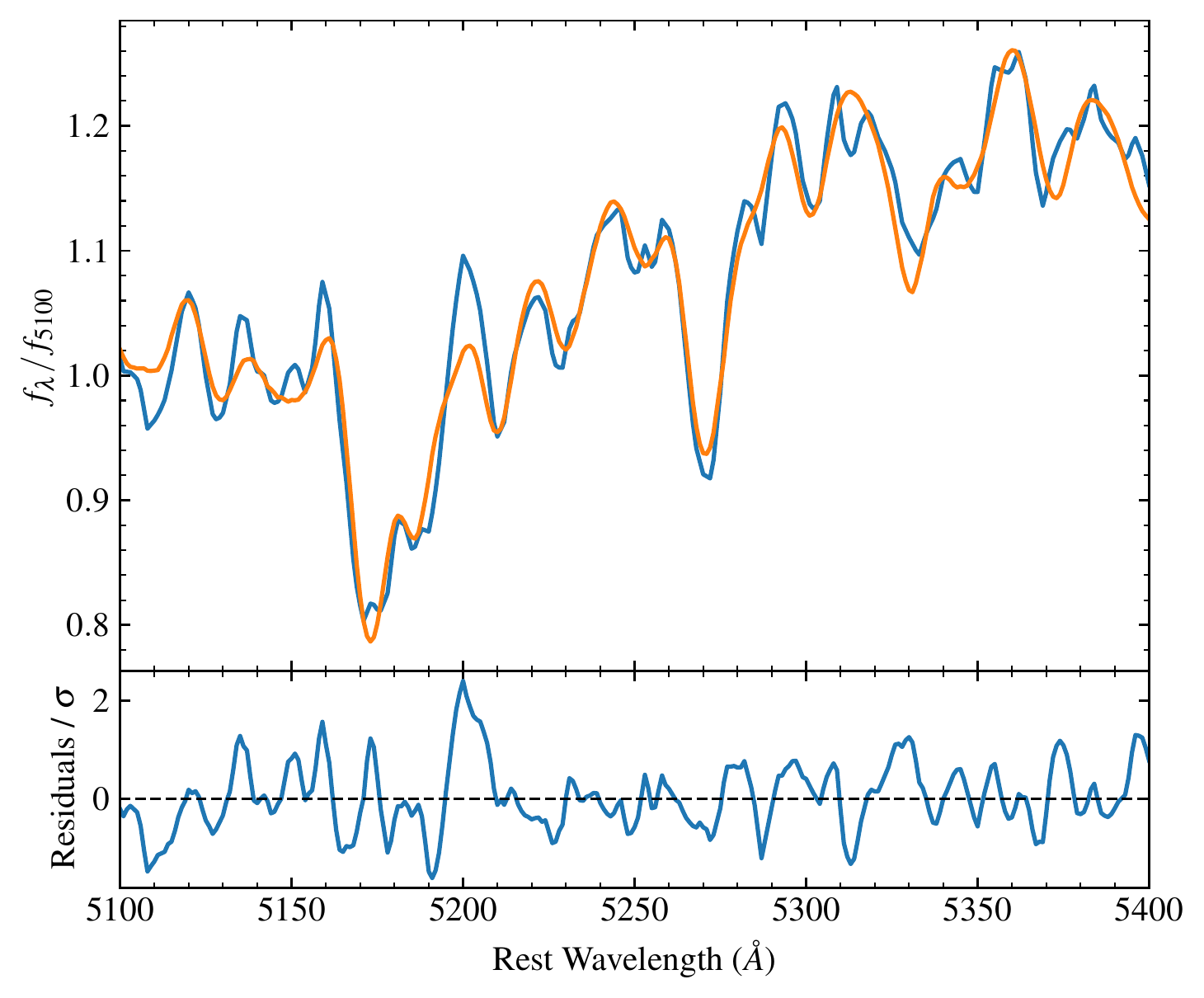}}
\caption{Upper panel: Best fit obtained with the {\sc starlight}-code to the \textit{2.1m/B\&Ch} spectrum of \twomas. The residuals show the clean \twomas\, spectrum. Lower panel: A second spectral fit focusing on a small region of the spectrum which was used to estimate the systemic velocity of \twomas.} 
\label{Fig:2mashost}
\end{center}
\end{figure}

\textsc{starlight} works by representing the observed spectrum as a linear combination of spectra in a given base, which are later convolved with a Gaussian kernel to fit the kinematics, at the same time accounting for the dust in the target galaxy by simultaneously fitting an attenuation curve.
In our case, the base was composed by simple stellar populations (SSP) from the MILES library \citep{2010MNRAS.404.1639V}.
For regions affected by atmospheric absorption and by H$\alpha$+[NII] and H$\beta$+[OIII] emission line complexes, different masks were used in order to exclude these features from the fit. The fitted spectrum after one hundred realizations is presented in Figure~\ref{Fig:2mashost}. 

In order to estimate the kinematic parameters, we used a smaller portion of the spectrum with numerous stellar absorption lines, between 5100{\AA} and 5400{\AA}. The uncertainties in the best fit parameters were derived from 100 Monte Carlo realizations, each time having the flux in each pixel randomized by a normal distribution with a standard deviation equivalent to a signal-to-noise ratio of 32. The obtained systemic velocity is V$\rm{_{R}}$\,=\,10057$\pm$18 km\,s$\rm{^{-1}}$.
Therefore,  \twomas\, is the companion galaxy of \iras\, since the latter has  a systemic velocity of V$\rm{_{R}}$\,=\,9893.1525$\pm$0 km\,s$\rm{^{-1}}$. We estimate a stellar velocity dispersion of $\sigma_{\star}$\,=\,191$\pm$10\,{\rm km\,s$^{-1}$}, based on the best fit broadening of $\sigma$\,=\,227$\pm$10\,{\rm km\,s$^{-1}$}, an instrumental broadening of 137$\pm$31\,{\rm km\,s$^{-1}$}, and the 58.4\,km\,s$^{-1}$ resolution of the MILES spectral library.

Once the contribution of the host galaxy was subtracted, the fit was done without the need of a power law. The pure emission lines were modelled using the \textsc{Specfit} code. Gaussian components were used to obtain the best fit, see Figure~\ref{Fig:2mashost} and Table~\ref{tab:fit-twomas}. The pure emission of \twomas\, shows that this object is an obscured Type 2 (Sy2) AGN \citep{1993ARA&A..31..473A}.


\begin{table*}
\caption{\bf Profile modelling results of \twomas}
\label{table:twomasmodel}
\begin{tabular}{lcccccc}
\hline \hline \\
\textit{2.1m/B\&Ch} & Flux\,$\times$10$^{-14}$ & Luminosity\,$\times$10$^{41}$ & $\lambda_{cent}$   & FWHM \\
              &  erg\,s$^{-1}$\,cm$^{-2}$\,\AA$^{-1}$ & erg\,s$^{-1}$ & \AA  &\kms \\
\hline \\
\hb      	&	13.41	$\pm$	0.67	&	3.36	$\pm$	0.08	&	4863.4	$\pm$	0.3	&	641	$\pm$	9	\\
\oiiib      	&	70.70	$\pm$	1.01	&	17.69	$\pm$	0.13	&	5009.1	$\pm$	0.1	&	641	$\pm$	9	\\
\ha      	&	79.01	$\pm$	1.85	&	19.77	$\pm$	0.23	&	6565.2	$\pm$	0.2	&	601	$\pm$	10		\\
\niib      	&	69.63	$\pm$	1.63	&	17.42	$\pm$	0.20	&	6585.4	$\pm$	0.2	&	601	$\pm$	10	\\
\siia     	&	41.53	$\pm$	1.68	&	10.39	$\pm$	0.21	&	6717.8	$\pm$	0.3	&	601	$\pm$	10	\\
\siib      	&	30.27	$\pm$	1.53	&	7.57	$\pm$	0.19	&	6734.4	$\pm$	0.5	&	601	$\pm$	10	\\
\oia      	&	20.83	$\pm$	1.24	&	5.21	$\pm$	0.15	&	6301.2	$\pm$	0.4	&	601	$\pm$	10	\\
\hline \\
\end{tabular} 
\label{tab:fit-twomas}
\end{table*}
\subsection{Black hole mass estimates}

Previous spectroscopic studies on \iras\, have obtained \mbh\, estimates using the complete broad component \citep[e.g.][]{2010ApJ...710..503W,2012MNRAS.419.2529R}. Nevertheless, due to the complex broad profiles shown by the Balmer lines of \iras\, (e.g. in Figure ~\ref{Fig:all-hbeta}) an estimation of the black hole mass M$\rm{_{BH}}$ can be obtained  using an empirical correction applied to this type of lines \citep[see][]{Marziani2019}. In particular, this correction is applied if the VBC of \iras\, is produced by a radial velocity excess that could be associated with gas with a velocity component increasing towards the central black hole. So, in that case, the full profile was symmetrized through the relation FWHM$_{symm}$\,=\,FWHM\,$-\,2\,c(\frac{1}{2})$, where c$(\frac{1}{2})$ is the velocity centroid estimated at half-maximum. So, assuming that the FWHM$_{symm}$ is a virial estimator, the \mbh\, of \iras\, is log\,M$\rm_{BHsymm}$\,=\,8.59\,$\pm$\,0.14 (M$_\odot$). Another way to correct the complete FWHM empirically is to consider the spectral type associated with the parameter space of Eigenvector 1, under the assumption that the broad symmetric component of \hb\ is representative of the virialized region of the BLR. \iras\, is a population B AGN in this parameter space, since its FWHM(\hb)\,$>$\,4000\,\kms. In this case, the correction proposed by \citet{Marziani2019} is FWHM$_{vir}$\,=\,0.8\,FWHM, and the obtained black hole mass estimate is log\,M$\rm_{BHvir}$\,=\,8.44$\pm$0.13 (M$_\odot$). Both results are in agreement within a factor of $\sim$1.4. In a recent work by \citet{2020MNRAS.492.4216S} an estimation of the M$_{BH}$ of \iras\, is log\,M$\rm_{BH}$\,=\,8.69\,$^{+0.22}_{-0.17}$ (M$_\odot$). This result is in agreement with our values, although the authors do not specify the method used in their estimation. The bolometric luminosity was computed using the luminosity continuum at 5100\AA, and a bolometric correction factor of 10.33 \citep{richardsetal06}. We obtained a value of $\log$ \lbol(IRAS) = 45.11 $\pm$ 0.09 (erg s$^{-1}$). Table \ref{table:bhm} shows the results of the black hole mass estimations for all the cases. With these results, the Eddington ratio \eddr\ = \lbol/\ledd\, was computed for \iras, assuming an Eddington luminosity of \ledd\ = 1.5$\times$10$^{38}$ (M$_{BH}$/M$_\odot$) erg s$^{-1}$. The Eddington ratios for the two \mbh\ estimations obtained for \iras\, are \eddr$_{(symm)}$ = 0.026 $\pm$ 0.020 and \eddr$_{(vir)}$ = 0.037 $\pm$ 0.020. 

In the case of \twomas, since it is a Type 2 object, the \mbh\, was obtained using the velocity dispersion $\sigma_{\star}$ of bulge stars obtained in Section~\ref{model2} and the M$\rm{_{BH}}$ vs. $\rm{\sigma_{\star}}$ relation from \citet{2013ApJ...764..184M}. Therefore, $\log$\,M$\rm{_{BH}}$\,=\,8.21$\pm$0.2 (M$_\odot$). The bolometric luminosity was computed using the \oiiib\ luminosity corrected by the Balmer decrement  L$^c_{\rm [OIII]}$, given by \lbol\ = C$_{\rm [OIII]}$ L$^c_{\rm [OIII]}$ with a bolometric correction factor C$_{\rm [OIII]}$ = 142 \citep{Lamastra2009}. We obtained for \twomas\, a bolometric luminosity $\log$\,\lbol  = 45.26 $\pm$ 0.04  (erg s$^{-1}$), and an Eddington ratio \eddr\ = 0.023 $\pm$ 0.003.

\begin{table*}
\caption{\bf Black hole mass estimates}
\label{table:bhm}
\begin{tabular}{lcccccc}
\hline \hline \\
 Object/method & FWHM & log \mbh & \eddr \\
 &  \kms & M$_\odot$ & \\
\hline \\
\iras\\
Sym  & 6549 $\pm$ 403 & 8.59 $\pm$ 0.14 & 0.026 $\pm$ 0.020\\
Vir  & 5526 $\pm$ 314 & 8.44 $\pm$ 0.13 & 0.037 $\pm$ 0.020\\
\hline \\
\twomas\\ 
M-$\sigma_\star$ & - & 8.21 $\pm$ 0.20 & 0.023 $\pm$ 0.003\\
\hline \\
\end{tabular} 
\end{table*}

\subsection{BPT Diagrams}
\label{BPT}

BPT diagnostic diagrams \citep[][]{1981PASP...93....5B,1987ApJS...63..295V} were used to classify \iras\, and \twomas. In all BPT diagrams shown in Figure~\ref{Fig:BPT} the solid curved line traces the upper theoretical limit due to pure HII region contribution established by \citep[][]{2001ApJ...556..121K}. The dotted line in the [NII] diagnostic provides an empirical upper limit,  measured by \citet{2003MNRAS.346.1055K}. The region lying between these two lines represents objects with spectra showing emission from HII regions, but also from the AGN. Sy galaxies occupy the upper region of the diagrams, and the dotted-lines in the [SII] and [OI] diagrams show the region of sources with LINER-like emission \citep[][]{2006MNRAS.372..961K}. The line ratios, obtained with the more intense narrow component, are presented in Table~\ref{tab:BPT}. In the three diagrams, \iras\, and \twomas\, appear in the AGN region. 

\begin{figure}
\begin{center}
{\includegraphics[width= 9cm]{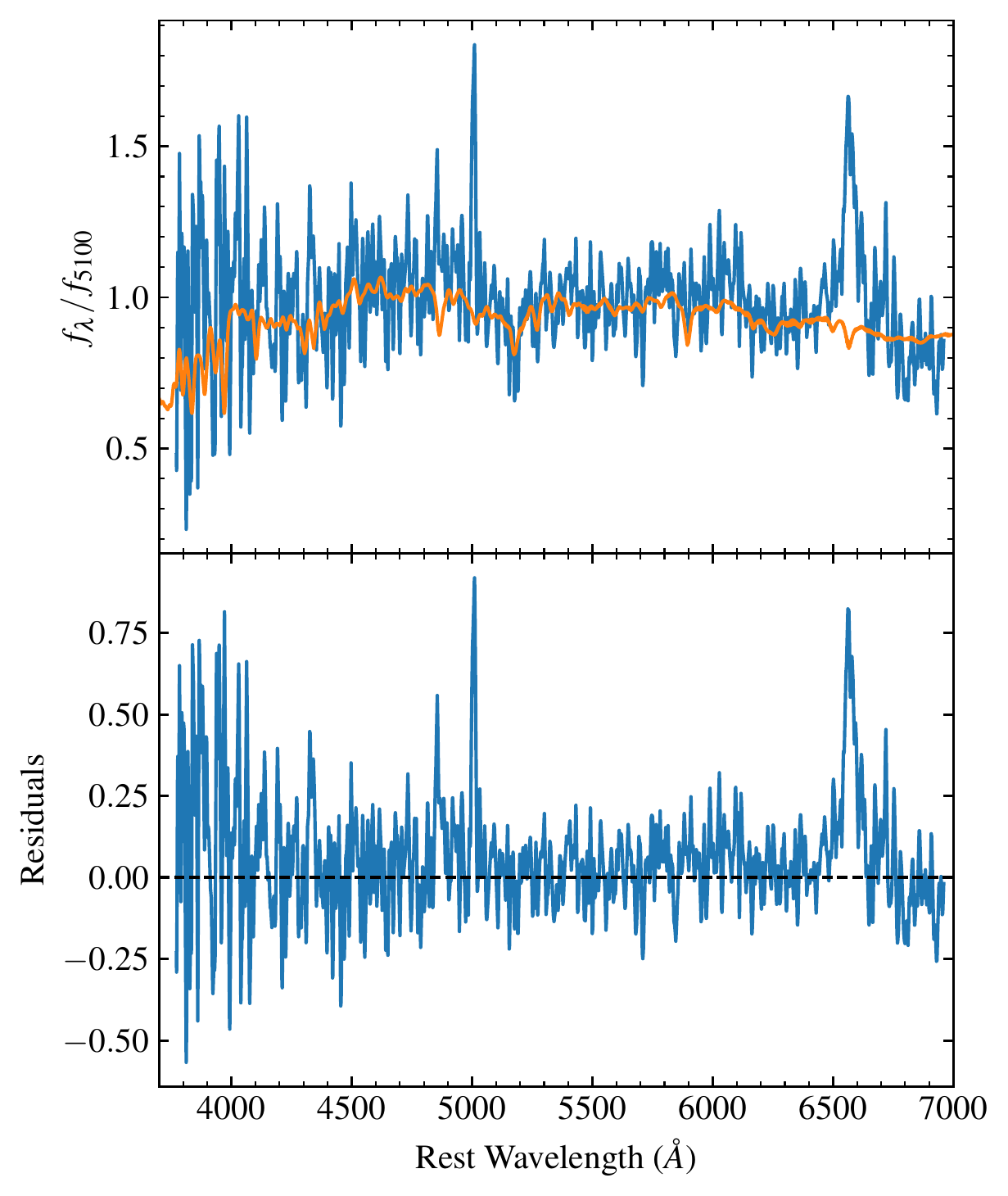}}
\caption{Best fit obtained with \textsc{STARLIGHT} of the bridge zone spectrum of \iras. After removing the host galaxy  contribution, the line ratios were estimated.} 
\label{Fig:bridge-SPM}
\end{center}
\end{figure}

\begin{figure*}
\subfigure{\includegraphics[width=16cm]{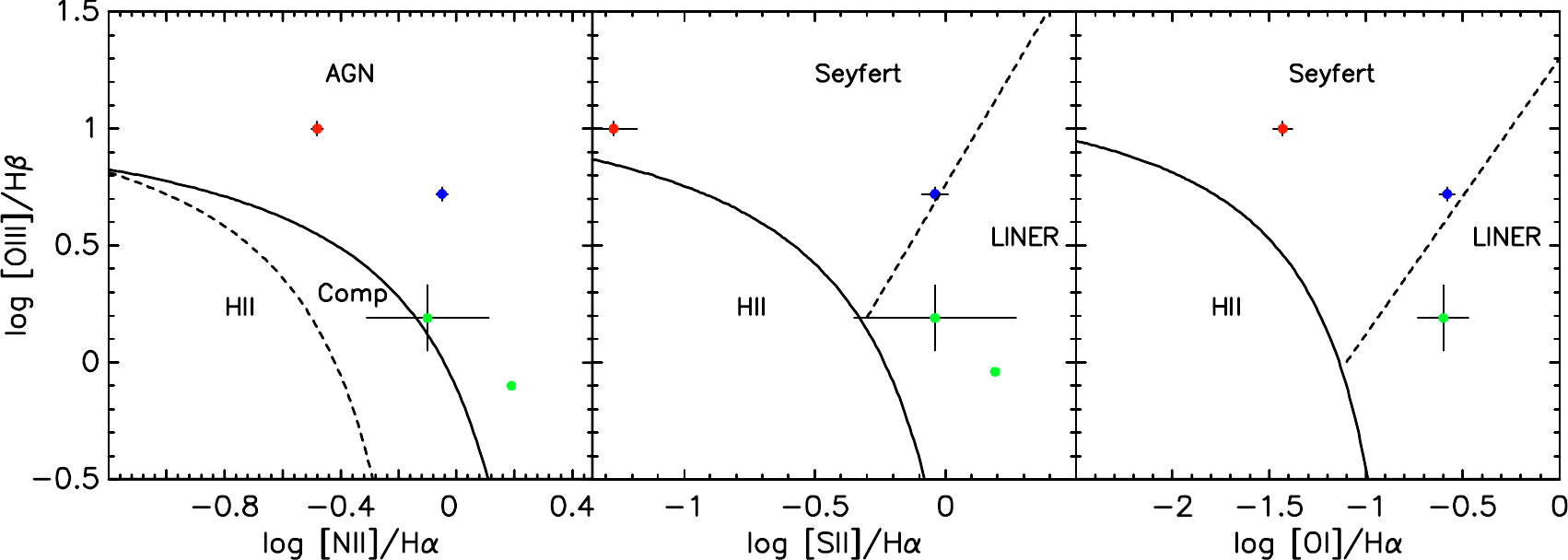}}
\caption{BPT diagnostic diagrams showing the AGN nature of \iras\,(red) and \twomas\,(blue). Since \iras\, has a double narrow line component, the more intense narrow component was used to estimate the line ratios. Also, the bridge zone is shown (green). The bridge zone appears in the LINER region.}
\label{Fig:BPT}
\end{figure*}

In order to analyse the spectrum of the bridge zone, the \textsc{starlight} code was used following the methodology explained in Section \ref{model2}. After fitting and subtracting the contribution from the host galaxy, line ratios were estimated from the emission spectrum (see Figure~\ref{Fig:bridge-SPM}) and are shown in green colour in the BPT diagrams. In all diagrams, the bridge zone emission appears in the LINER-like region.

\begin{table*}
\caption{\bf BPT line ratios}
\begin{tabular}{lcccc}
\hline \hline \\
Line ratios & 2MASX J06021107+2828382   & IRAS\,05589+2828 &  Bridge zone\\
\hline \\
log\oiiib/\hbnc	& 0.72$\pm$0.03	&	1.00$\pm$0.03 	&	0.19$\pm$0.14	\\
log\niib/\hanc	& -0.05$\pm$0.02 &	-0.48$\pm$0.02 	&	-0.10$\pm$0.21	\\
log\sii/\hanc	& -0.04$\pm$0.05 &	-1.27$\pm$0.09 	&	-0.04$\pm$0.31	\\
log\oia/\hanc   & -0.58$\pm$0.04 &	-1.43$\pm$0.05 	&	-0.60$\pm$0.13	\\
\hline
\end{tabular}
\label{tab:BPT}
\end{table*}

\section{\textit{VLA}--data}
\label{radio}

\begin{figure*}
\begin{center}
\includegraphics[width=0.64\textwidth]{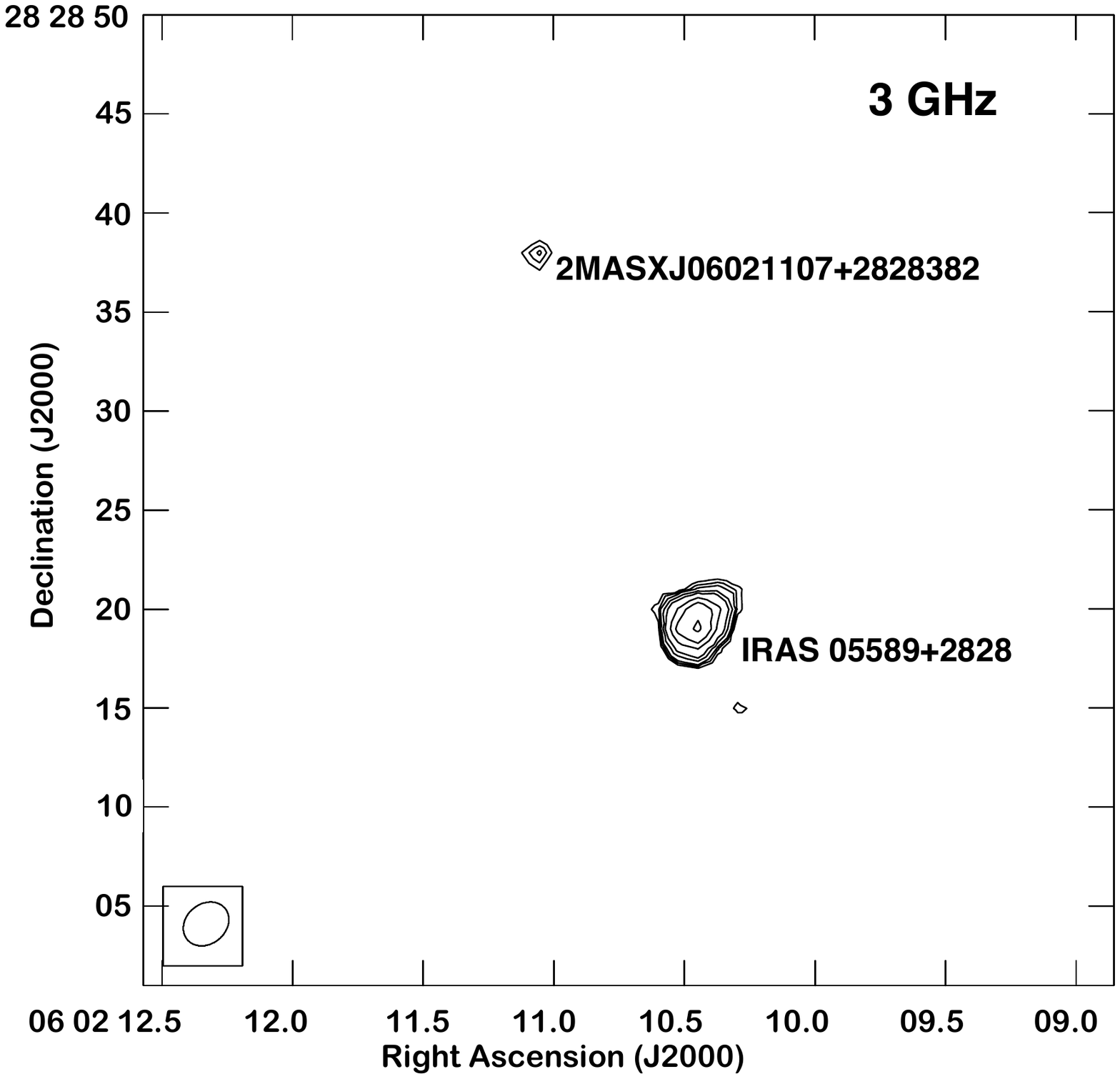}
\includegraphics[width=0.5\textwidth]{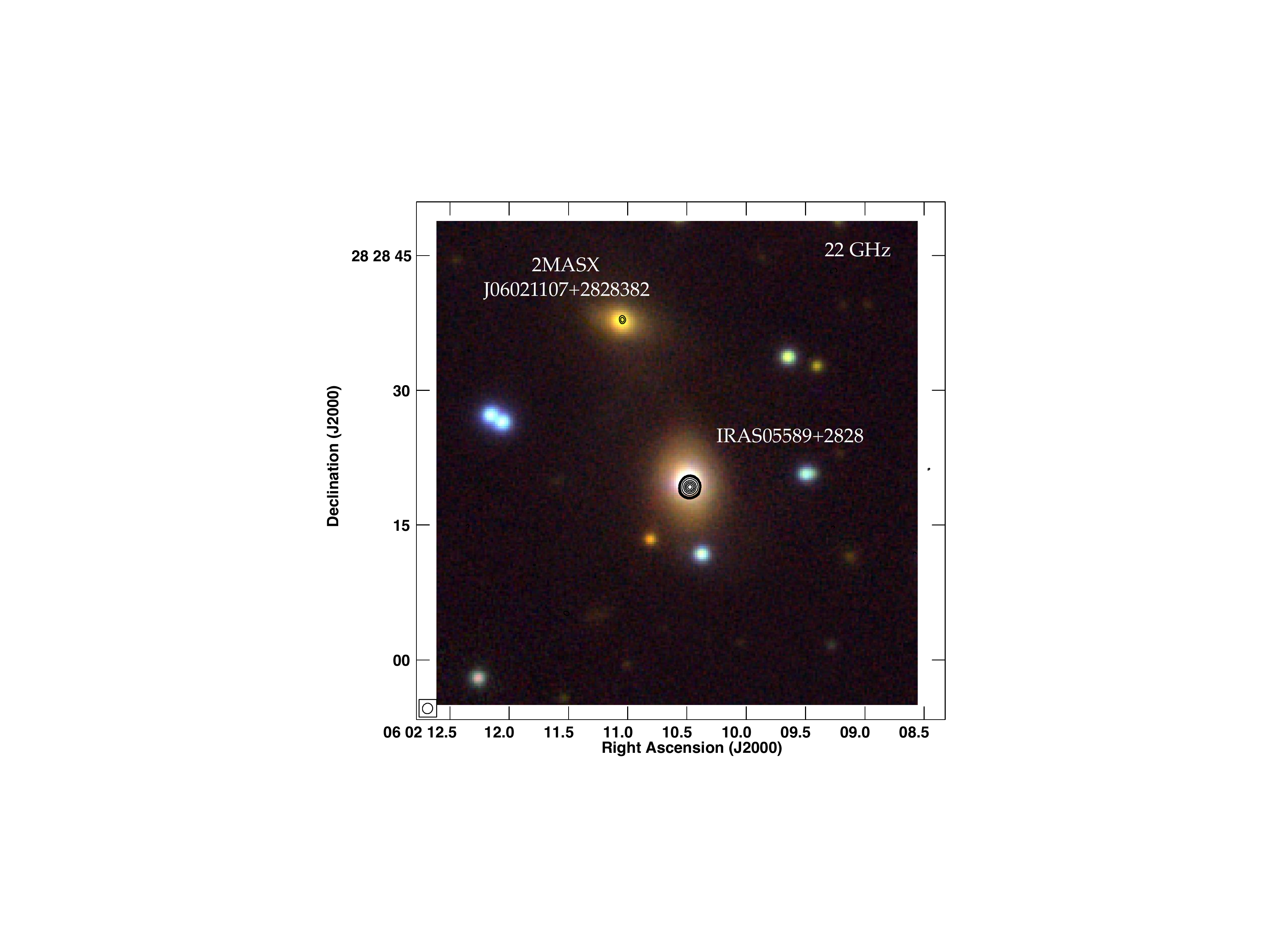}
\caption{Upper panel: Contour image of the \iras\, region at 3 GHz. Contours are -4, 4, 5, 6, 8, 10, 15, 20 and 30 times 130 $\mu$Jy~beam$^{-1}$.  The sources  \iras\, and 2MASXJ06021107+2828382 are indicated. The beam 
($2\rlap.{''}47 \times 2 \rlap.{''}01$; PA = $-51^\circ$) is shown in the bottom left corner. Lower panel: Contour image of the \iras\,  region at 22 GHz with PanSTARRS/DR1 composite z and g bands image. Contours are -4, 4, 5, 6, 8, 10, 20, 40, 60, 80 and 100 times 26 
$\mu$Jy~beam$^{-1}$. The sources \iras\, and 2MASXJ06021107+2828382 are indicated. The beam 
($1\rlap.{''}18 \times 1 \rlap.{''}12$; PA = $-32^\circ$) is shown in the bottom left corner. The image has been corrected for the
response of the primary beam.}
\label{Fig:VLA}
\end{center}
\end{figure*}

In the archives of the Karl G. Jansky Very Large Array (VLA) of NRAO\footnote{The National Radio Astronomy Observatory is a facility of the National Science Foundation operated
under cooperative agreement by Associated Universities, Inc.} observations of \iras\, were found as part of the project 13A-281 during 2013 July 20. These observations were made in the radio K-band (18 to 26 GHz, with a central value of 22\,GHz) with the VLA in the C configuration. The data were calibrated in the standard manner using the CASA (Common Astronomy Software Applications) package of NRAO and the pipeline provided for VLA\footnote{https://science.nrao.edu/facilities/vla/data-processing/pipeline} observations. This pipeline consists of a sequence of individual tasks that are applied to the data with the final goal of calibrating it in amplitude and phase. Images are then obtained by Fourier transforming of the calibrated data. In addition, the data were self-calibrated in phase once. Additional self-calibration iterations did not produce any significant improvement.  These observations have been reported previously by \citet{2020MNRAS.492.4216S}. They made images with robust weighting of -0.5 \citep{1995AAS...18711202B}, 
to compromise between sensitivity and angular resolution, while we used a robust weighting of 2, to obtain the highest sensitivity possible. \citet{2020MNRAS.492.4216S} detect \iras\, with a total flux density of 2.79 mJy, while we do it with a total flux density of 3.0$\pm$0.1 mJy (see left contour image in Figure~\ref{Fig:VLA}). The slightly larger value determined by us from the same data is probably due to the different weighting schemes.

In addition to \iras, we report for the first time the detection at 22 GHz of 2MASXJ06021107+2828382 with a total flux density of 0.20$\pm$0.03 mJy (see Figure~\ref{Fig:VLA}). These two sources are the only ones above 5\,$\sigma$ in a region of 1\arcmin$\times$1\arcmin centered on \iras\, (Figure~\ref{Fig:VLA}). Both sources appear unresolved for the angular resolution of
$\sim$1\arcsec. In Table~\ref{table:VLAdata} we summarize the parameters of the two sources.

 \begin{table*}
\caption{\bf Parameters of the VLA sources.}
\label{table:VLAdata}
\begin{tabular}{lcccccc}
\hline\hline
    Source                &  \multicolumn{2}{c}{Position$^a$  }      &      \multicolumn{2}{c}{Flux Density (mJy)}      &        Spectral  \\
&    RA (2000)   &    DEC (2000)  &   3 GHz      &      22 GHz    &      Index \\ 
\hline
\iras\,     &         06:02:10.475$\pm$0.001 &     28:28:19.24$\pm$0.01   &        5.0$\pm$0.3      &     3.0$\pm$0.1    &         -0.26$\pm$0.03 \\
2MASXJ06021107+2828382    &    06:02:11.029$\pm$0.032  &    28:28:37.94$\pm$0.37  &         0.81$\pm$ 0.14  &    0.20$\pm$0.03      &   -0.70$\pm$0.11 \\
\hline
\multicolumn{4}{l}{$^{a}$ Positions from the 22 GHz data.}\\
\end{tabular}\\
\end{table*}

To determine the spectral index of these sources, a {\it Quick Look} image from the first epoch of the Karl G. Jansky Very Large Array Sky Survey (VLASS) was used.
A detailed description of this survey is given by \citet{2020PASP..132c5001L}. 
In brief, the observations of this survey are made in the S-band (2 to 4 GHz, with a centre value of 3 GHz) with a typical angular resolution of $2\rlap.{''}5$.
The {\it Quick Look} images of the VLASS have limitations, in particular the positions are accurate only to $\sim 0\rlap.{''}5$ and the 
total flux densities of the second campaign of the first epoch (when the image of interest was obtained) 
are systematically low by 3\%. We corrected the image for this underestimate.
A more detailed discussion of the limitations is given in the homepage of 
NRAO. \footnote{\url{https://science.nrao.edu/vlass/data-access/vlass-epoch-1-quick-look-users-guide}}

The contour images at 3 GHz and 22 GHz from the VLASS are presented in the upper and lower panels in Figure~\ref{Fig:VLA}, respectively. As in the case of the 22 GHz image,
the only sources detectable above 5-$\sigma$ in the region are \iras\,  and 2MASXJ06021107+2828382. The compactness of both sources favours an AGN nature.

The flux densities are given in Table~\ref{table:VLAdata}, as well as the spectral indices derived from the 22 and 3 GHz observations. The spectral index of \iras\, of -0.26$\pm$0.03 is flat, which is characteristic of compact AGN, while the spectral index of 2MASXJ06021107+2828382 of -0.70$\pm$0.11 favours optically thin synchrotron emission from an AGN \citep[see][]{2015ApJ...813..103M}. The separation between the two sources is 20.08\arcsec\, which at the distance of \iras\, corresponds to $\sim$13.3\,kpc.


\section{\textit{X-ray}--data}
\label{X-ray}

IRAS\,05589+2828 has been pointed in X-ray with both \swift\, and \chandra\, observatories. All observations included both IRAS\,05589+2828 and its companion galaxy \twomas. The sources were in the field of view of a series of three \swift\, observations with identification numbers
3525501, 3525502, and 3525503 with respective exposures of $\sim$7.5, 5.1, and 4.3\,ks, and observed during 2006 March 3 and 6, and April 21, respectively. The \chandra\, observation, with identification number  12864, was obtained on 2011 January 16 with a total exposure time of $\sim$16\,ks. At the moment, {\it XMM-Newton} has not pointed IRAS\,05589+2828, but the source was detected in the second version of the \textit{XMM-Newton Slew-Catalog} \citep[][]{2008A&A...480..611S} from an observation during 2003 March 26, reporting a count rate of 3.96\,counts s$^{-1}$. The associated estimated flux in the 0.2\,-\,12\,keV band is 1.3 $\pm$ 0.3$\times$10$^{-11}$\,erg\,cm$^{-2}$\,s$^{-1}$. In the following, we will analyse the \swift\, and \chandra\, data available, studying both the images and spectroscopic data.

For both \swift\, and \chandra\,data, we have followed the standard procedures for their reduction described in the respective satellite documentation. For the \swift\, observations, we have only considered the {\it XRT} data. The three available observations have been obtained with the {\it Photon Counting} mode, using the 600$\times$600 pix$^{2}$ (23.6$\times$23.6 \arcmin$^{2}$) pointing configuration. The data were reduced according to the procedure described in {\it The SWIFT XRT Data Reduction Guide v1.2}. \chandra\, observed the source using the {\it ACIS-S} detector; the data have been reprocessed according to the standard procedure and using the {\it CIAO v4.13}\footnote{See \url{https://cxc.cfa.harvard.edu/ciao/} and \cite{fruscione}} and the {\it CALDB 4.9.4} . 

X-ray spectra of all available observations have been extracted and analysed.
For the three \swift\, observations of the whole X-ray emission inside a circle of radius 1\arcmin\,  were extracted. The background was extracted in a region of the same size located close to the source but free of any spurious emission. The \swift\, spectra have been then grouped in order to have at least 15 counts per bin. In order to extract and analyse the \chandra\, spectrum of the source, we have extracted the source and background spectra using the reprocessed \chandra\, data, a level\,=\,2 event file with the standard filtering. The source spectrum was obtained from a circular region of 7.6\arcsec, centred in the pixel of maximum emission of the source. The background spectrum was obtained from the combination of four circular regions of 9\arcsec\,located close to IRAS\,05589+2828 but free from any other source of emission. The spectrum has then been grouped in order they have at least 15 counts per bin in order to be able to apply the $\chi^2$ statistics. The spectral analysis has been performed using the package {\it sherpa} inside {\it CIAO\,v4.11}\footnote{See \url{https://cxc.cfa.harvard.edu/sherpa/} and \cite{sherpa}}.

\subsection{X-ray Spatial Analysis}
\label{x-raysp}

\begin{figure*}
\begin{center}
\includegraphics[width=0.75\textwidth]{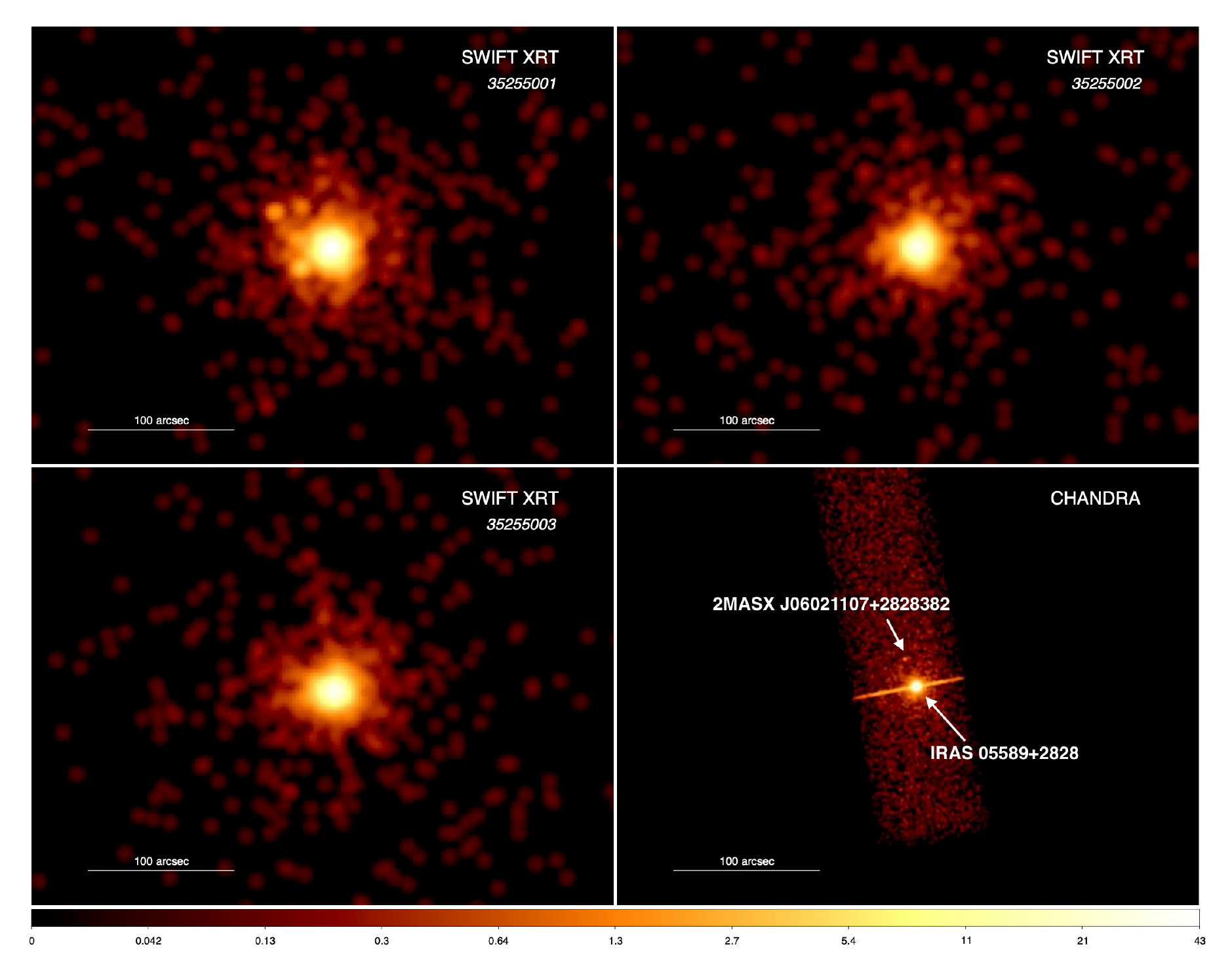}
\caption{From left to right and top to bottom, the smoothed images of the three available observations of {\it XRT} and the \chandra. All images have the same field of view and orientation, North is up and East is left, for comparison purposes. The \chandra\, spatial resolution allows detecting a point-like source towards the North of \iras\, that, it is actually identified with the galaxy \twomas.}
\label{fig:swiftchandra}
\end{center}
\end{figure*}

\begin{figure}
\begin{center}
\includegraphics[width=\columnwidth]{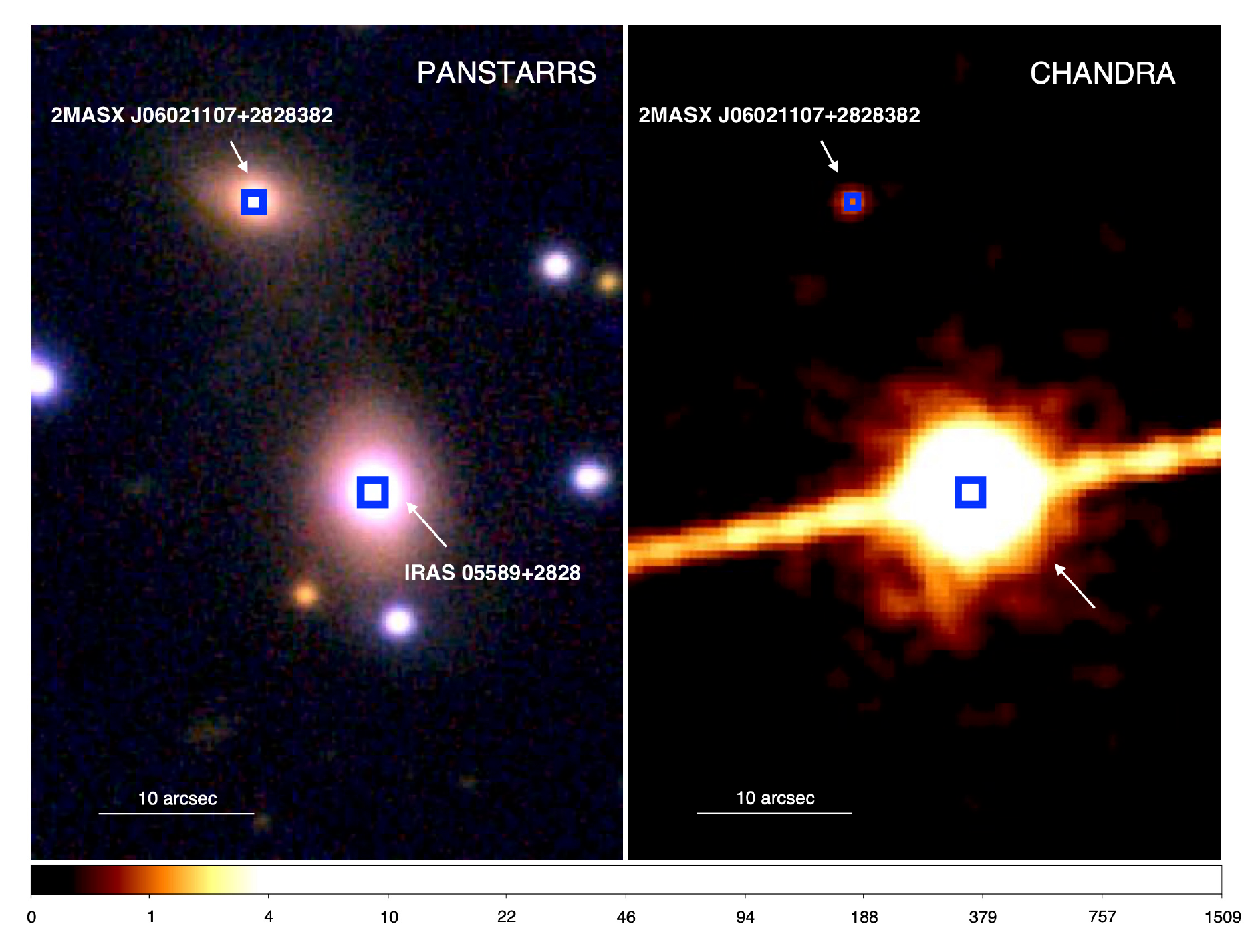}
\caption{Left panel: Field of view of {\it PanSTARRS} obtained with the {\it z} and {\it g} filters. Right panel: \chandra\, image showing the same field of view for comparison. The maximum X-ray emission of the two point-like sources has been marked in both images with a blue box. The location of \iras\, and \twomas\, in {\it PanSTARRS} coincides with X-ray point-like sources.}
\label{fig:panschandra}
\end{center}
\end{figure}

For the spatial analysis of the IRAS\,05589+2828 field of view, we have considered the three {\it XRT} \swift\, and the \chandra\, observations. Figure~\ref{fig:swiftchandra} shows the three \swift\, and \chandra\, images in the 0.5-10\,keV energy band. All four images have been smoothed. The images show that IRAS\,05589+2828 is detected as a point-like source in the \swift\, observations, but the PSF extends to the location \twomas. However, the \chandra\,resolution allows disentangling both \iras\, and a weak companion source. In Figure~\ref{fig:panschandra}, we compare the {\it PanSTARRS} image of source, on the left, with the \chandra\, image, on the right. The {\it PanSTARRS} image is a composition of the {\it z} and {\it g} filters. We have located the position of the maximum \chandra\, emission in both images, showing that the location of the brightest source coincides with the location of IRAS\,05589+2828 and the location of the weakest source coincides with the location of \twomas. No other sources have been detected in the X-ray images. It is worth noting that the  3525001 {\it XRT} observation, the deepest \swift\, observation, shows several point sources surrounding the East and North part of the X-ray bulk emission, as seen in the first panel of Figure~\ref{fig:swiftchandra}. These sources do not coincide with any of the sources detected in optical images. It is not clear if these sources are real or are only artefacts of the fluctuations of the X-ray images and not real X-ray sources. It is also worth noting that the \chandra\, image shows that IRAS\,05589+2828 clearly suffer from piled-up, last panel of Figure~\ref{fig:swiftchandra}, showing the characteristic signature of the saturation streak.

\begin{figure}
\begin{center}
\includegraphics[width=\columnwidth,height=0.75\columnwidth]{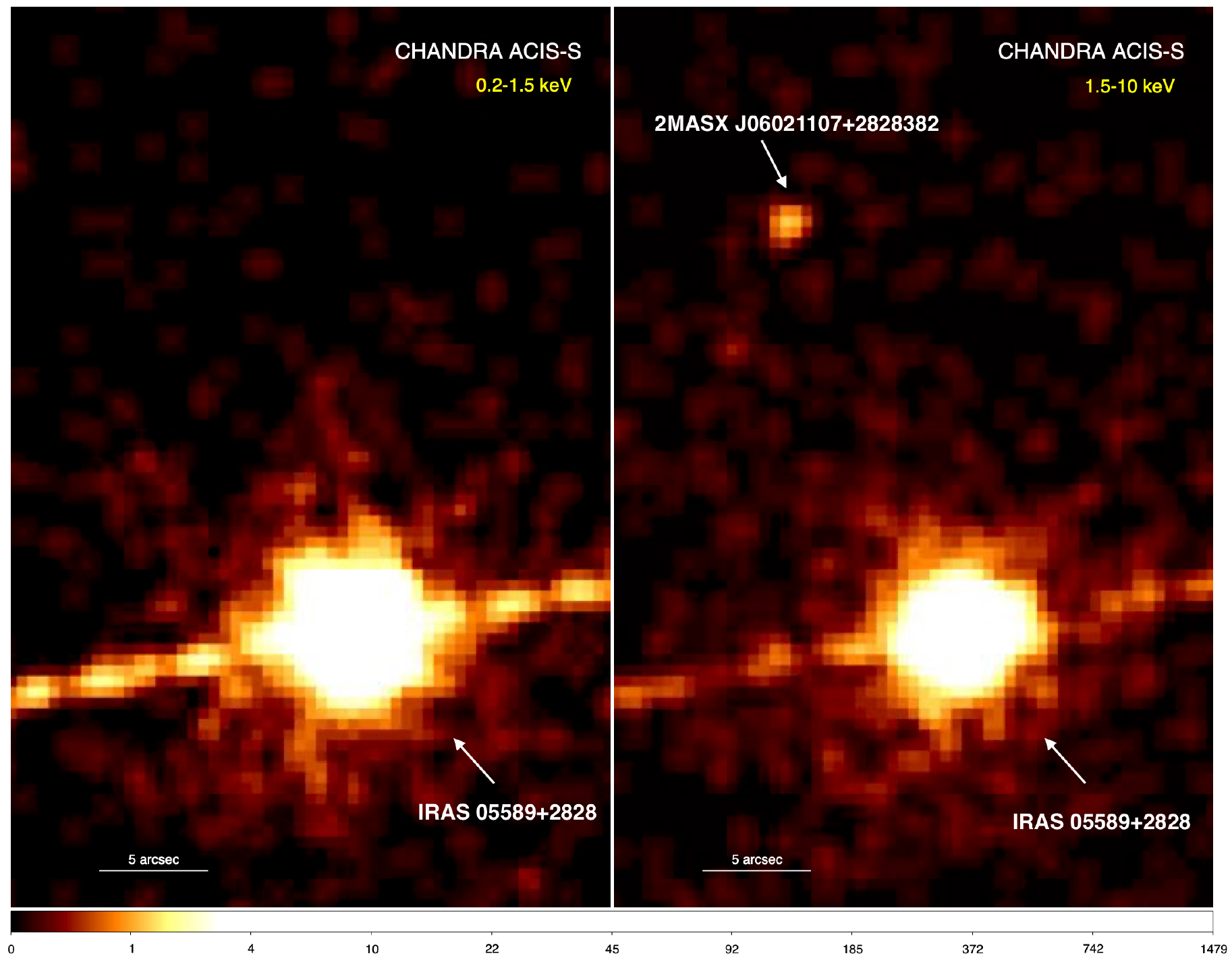}
\caption{\chandra\, images in the {\it soft} band (0.2-1.5\,keV) on the left, 
and {\it hard} band (1.5-10\,keV) on the right. The images show that \twomas\, 
is detected only in the hard band.}
\label{fig:chandrasofthard}
\end{center}
\end{figure}

In order to further explore the X-ray image, we have generated two \chandra\, images in the {\it soft} (0.2-1.5\,keV) and {\it hard} (1.5-10\,keV) bands. Both images can be seen in Figure~\ref{fig:chandrasofthard}. Interestingly, the weakest source, identified with \twomas, is not detected in the {\it soft} band. A close inspection shows that while in the hard band image, a total of 24 counts are detected in a circle of radius of 3 pixels, or 1.476\arcsec, only three counts are detected in the same extraction area, indicating that the source is likely to be highly absorbed, being the reason of the lack of photons at soft energies. This result is compatible with the Type 2 nature of \twomas, as found with our optical spectroscopic analysis.

\subsection{X-ray Spectral Analysis}
\label{xrayspec}

In order to study the spectra of \iras, we have tested several emission models for \chandra\ and \swift\ data. We have modelled all data sets individually, but also we have fitted all spectra simultaneously, in which only the normalization of the components for each of the spectra have been left unbounded among the observations. The results obtained for parameters and goodness of the fits are given in Table~\ref{Table:xrayfits}. A local galactic absorption, in the line of sight of \iras\, of 3.37$\times10^{21}$~cm$^{-2}$ \citep{2016A&A...594A.116H}, has been considered. 

For the \chandra\, observation, as it is shown in Figures~\ref{fig:panschandra} and \ref{fig:chandrasofthard}, the X-ray image shows a streak characteristic of the saturation of the CCD by the source. In order to account for pile-up, we have followed the {\it The Chandra ABC Guide to Pileup}. The pile-up was taken into account by including the {\it pileup} \citep{2001ApJ...562..575D} model when fitting the spectrum. The pile-up model has two free parameters: $\alpha$ which  represents the probability, per photon count greater than one, that the piled event is not rejected by the spacecraft software as a {\it bad event} and $f$, the fraction of events in the source extraction region to which {\it pileup} will be applied. For our data, the best fit values for each parameter are in the range of $\alpha\sim$0.40-0.70 and $f\sim$\,0.85\,-\,0.91, depending on the model. The pile-up fraction obtained for the several tested models ranged in 30\,-\,50\%. 

\begin{figure*}
\begin{center}
\includegraphics[width=12.7cm]{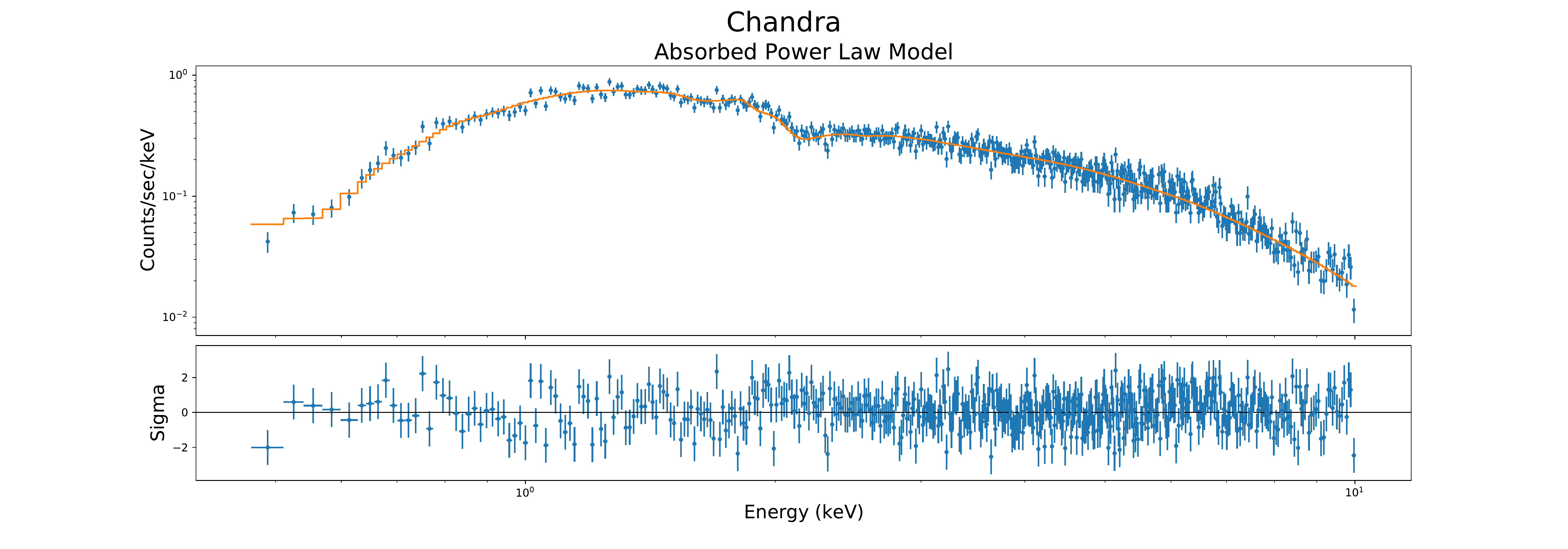}\\
\includegraphics[width=12.7cm]{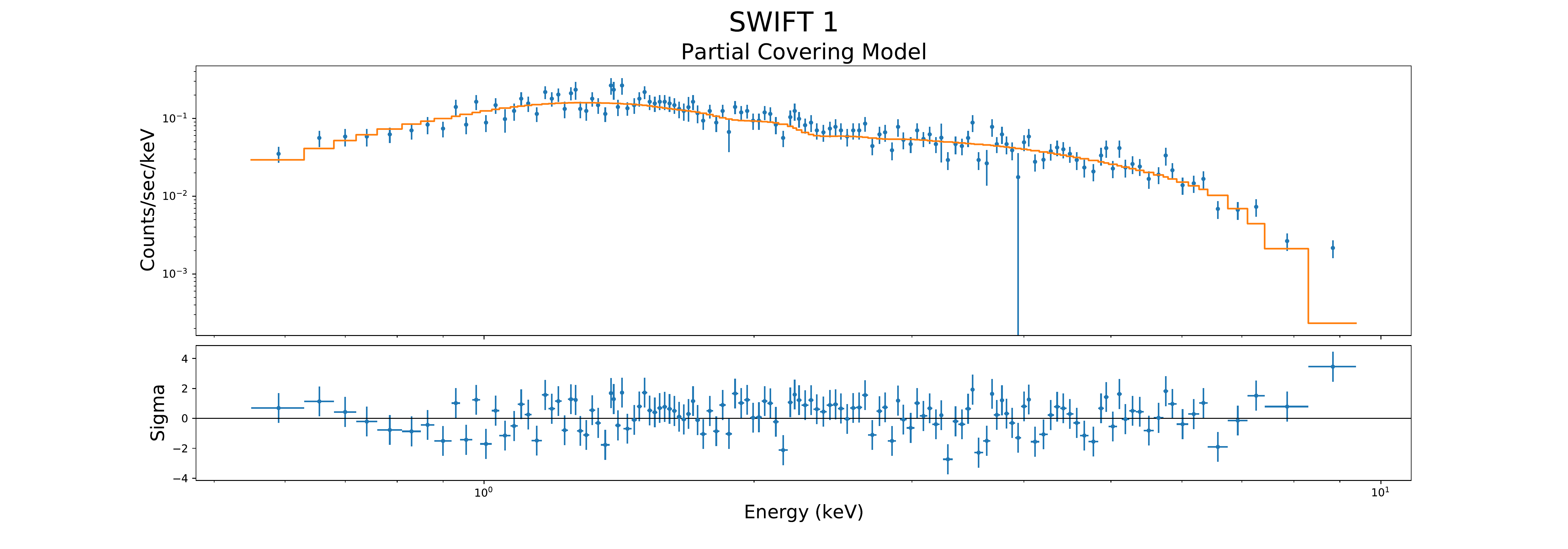}\\
\includegraphics[width=12.7cm]{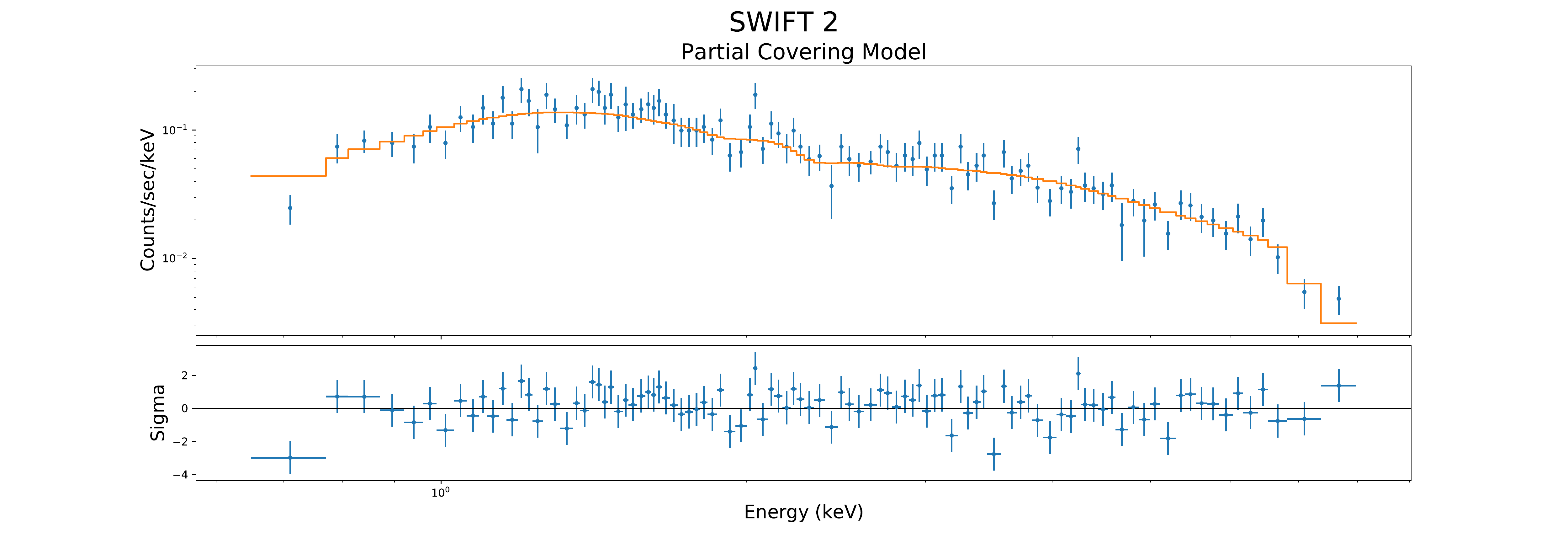}
\includegraphics[width=12.7cm]{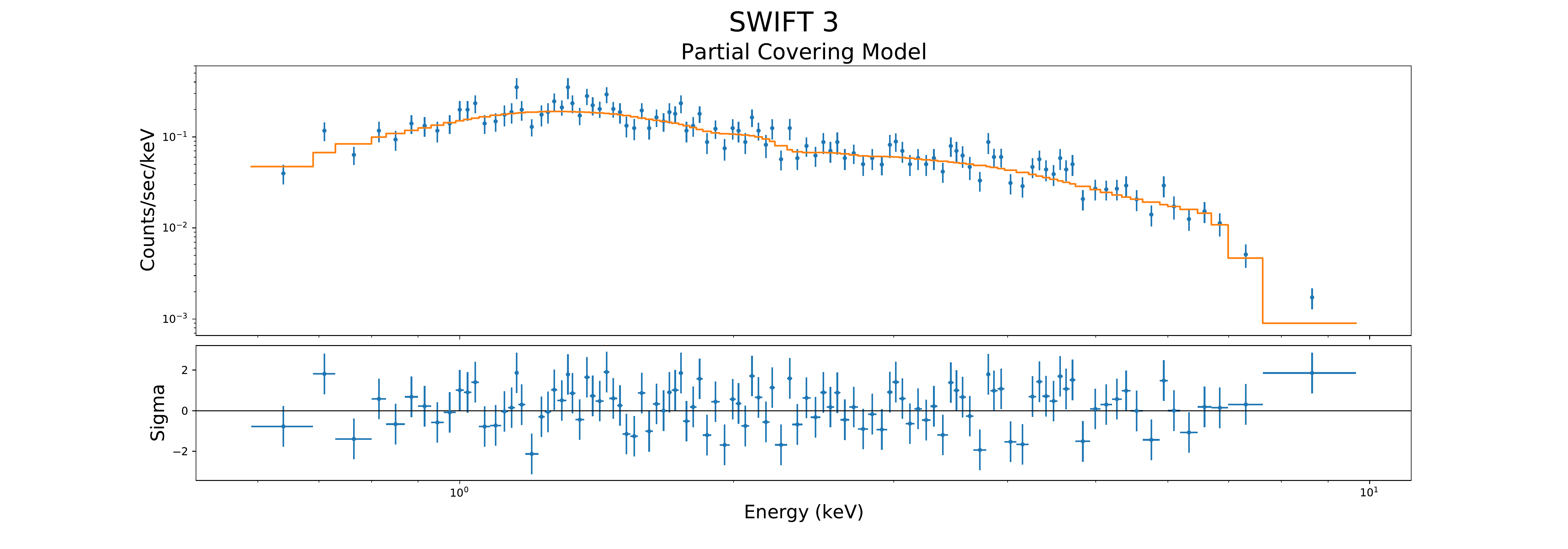}
\includegraphics[width=12.7cm]{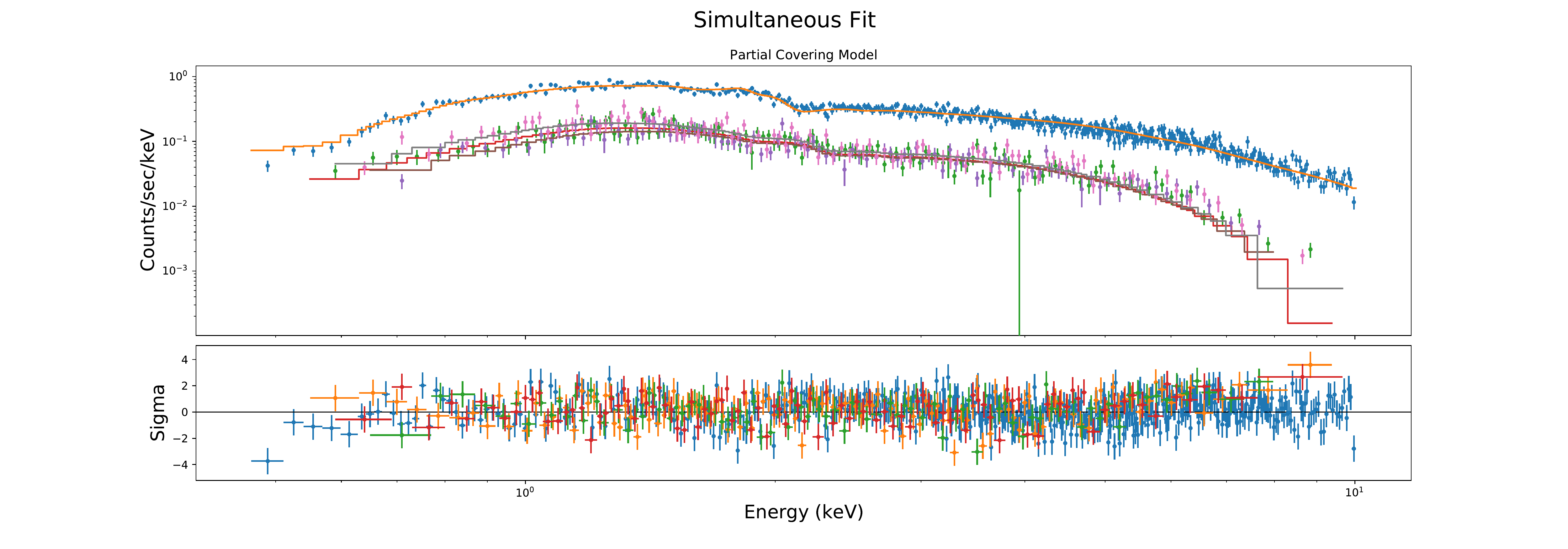}
\caption{From top to bottom, best model fits and residuals obtained for \chandra, SWIFT1-3 and all observed spectra of \iras. See Section \ref{xrayspec}\ and Table \ref{Table:xrayfits}\ for further details.}
\label{fig:swiftspectra}
\end{center}
\end{figure*}

The spectra shown in Figure~\ref{fig:swiftspectra} were fitted using: {\it i)} a single power law model, and {\it ii)} an absorbed power law at the redshift of the galaxy model. The {\it i)} single power law model provides a fairly acceptable description for all data sets with values of $\chi^2_{\nu}$ in the range of 1.06\,-\,1.24. The values of the power law index are in the range of 1.03 and 1.39, considered to be low for expected values of AGN sources.

We then considered if the  {\it ii)} absorbed power law model provides a better fit to our data sets. According to the F-test, only the fit for the \chandra\,data improves by the inclusion of a neutral absorption component. The resulted value for $\Gamma$ is $1.96\pm0.10$ and the N$_H$ is $1.7\pm0.3\times 10^{21}\,cm^{-2}$. The $\chi^2_{\nu}$ is 1.02 for 476 dof. However, for the \swift\,spectra and the simultaneous fit, the fits do not improve significantly and the values of the power law indices remained similar to the ones for the single power law model. At the same time, the values of the absorption obtained are low, of the order of $10^{20-21}\,cm^{-2}$. 

Motivated by the flatness of the power law and in order to further understand the X-ray emission of \iras, we also tested for all the spectra an {\it iii)} partial covering model. This model  provides a statistically better explanation for the simultaneous fit, with an F-test probability $>99.99$\%. The three \swift\, spectra fits also improve when this model is applied, with F-test probabilities of 96\%, 99.3\% and 99.56\% for \swift-1, 2, and 3 respectively. The intrinsic absorption values measured are in the range of $10^{23-24}\,cm^{-2}$, highly exceeding the local galactic extinction of 3.37$\times10^{21}$~cm$^{-2}$  \citep{2016A&A...594A.116H}. The indices of the power law remain flat, in the order of magnitude of $\Gamma=1.1-1.3$. When this model is applied to the \chandra\, spectrum there is no statistical improvement of the fit and, in fact, the found values are fully compatible with the absorbed power law model.

We have also tested the possibility of the presence of an iron line by adding this component to the best fit model of each spectrum. We have included a Gaussian line with the energy fixed to 6.4 keV and the energy width fixed to the instrumental resolution. For none of the data set, the inclusion of the Fe K${\alpha}$ line improves the fit. 

Table~\ref{Table:xrayfits}  also shows the derived fluxes and luminosities for the best fit model for each data set, marked in bold in the value of the $\chi^2_\nu$ in the same table. 

In the field of view of \chandra\, observation we have also detected \twomas. However, the counts of this source are not enough to extract a spectrum.  We have calculated the hardness ratio considering the 0.5\,-\,2\,keV and 2\,-\,10\,keV bands, obtaining a value of 0.85, with 25 counts in the hard band and only two counts in the soft band. Using the relationship in \citet{2006A&A...451..457T}, we have also estimated the flux in the 0.5\,-\,2\,keV and the 2\,-\,10\,keV bands, assuming a Galactic absorbing column of 8$\times10^{19}\,cm^{-2}$ and a photon index of $\Gamma$\,=\,1.4, the model which describes well the average spectrum of the sources detected in the Chandra Deep Field South. The values obtained are F$_{0.5-2\,keV}$\,=\,6.3$\times$10$^{-16}$\,erg\,s$^{-1}$\,cm$^{-2}$ and F$_{2-10\,keV}$\,=\,5.0$\times$10$^{-14}$\,erg\,s$^{-1}$\,cm$^{-2}$. The soft emission is negligible, probably due to a high intrinsic absorption \citep[see][]{2000ApJ...542..914W,2018ARA&A..56..625H}.



\begin{table*}
\caption{{\bf Results of the spectral analysis of the three \swift{\it XRT} and \chandra\, spectral analysis. Includes: count rate, values of the best fit single power-law model  parameters, fit goodness, fluxes and luminosities.}}
\label{Table:xrayfits}
\begin{tabular}{lllllll}
\hline\hline
Observation & Chandra &  ~SWIFT~1 & ~SWIFT~2 &  ~SWIFT~3 & Simultaneous \\
Obs. ID  & 35255001 &  35255002 & 35255003\\
\hline
Obs. Date & 03-03-2006 & 06-03-2006  &  21-04-2006 & 06-01-2011 \\
Exposure$^{a}$  & 6.4  & 5.0  & 4.3 & 14.5 \\
Count Rate$^{b}$ & 0.379$\pm0.08$	&	0.365$\pm0.08$	&	0.427$\pm0.10$ & 1.946$\pm$0.012 \\\hline\hline
\\
\underline {Model: phabs*powerlaw} \\

$\Gamma$$^{c}$ &  $ 1.39 \pm 0.03 $ & $ 1.16 \pm 0.04 $ & $ 1.03 \pm 0.05 $ & $ 1.23 \pm 0.05 $ & $ 1.240 \pm 0.019 $ \\

Normalization \chandra$^{e}$ &  $ 10.4_{ -1.7 }^{+ 1.4 }$ & & & & $14.2\pm1.1$\\
Normalization \swift1$^{e}$ & &  $ 3.14 \pm 0.12 $ & & &  $3.35 \pm 0.09 $ \\
Normalization \swift2$^{e}$ & & & $ 2.57 \pm 0.12$  & &  $3.06 \pm 0.09 $ \\
Normalization \swift3$^{e}$ & & & & $ 3.83 \pm 0.17$   &  $3.85 \pm 0.11 $ \\
alpha$^{f}$ & $ 0.51 _{ -0.03 }^{+ 0.05 }$ & & & & $ 0.435 _{ -0.009 }^{+ 0.01 }$\\
f$^{g}$ & $ 0.87 \pm 0.003 $ & & & & $ 0.867 _{ -0.002 }^{+ 0.003 }$\\
$ \chi^2$ & 537.27 & 159.38 & 104.94 & 124.19 &  974.97 \\
 DOF$^{h}$ & 477 & 129 &99 & 104&  812 \\
$ \chi^2_\nu$ & 1.13 & 1.24  & 1.06 & 1.19 &  1.20 \\

\underline {Model: phabs*zphabs*(powerlaw)}\\

$N_H$$^{d}$ &  $0.17\pm0.03$ & $<4\times10^{-2}$ & $0.10_{-0.06 }^{+0.07 }$ & $<3\times10^{-2}$ & $0.015\pm0.013$  \\

$\Gamma$$^{c}$ &  $ 1.96\pm0.10$ & $ 1.16_{ -0.04 }^{+ 0.06 }$ & $ 1.14\pm0.09$ & $ 1.23\pm0.05$ & $ 1.27\pm0.03$  \\
Normalization \chandra$^{e}$ &  $14\pm2$ & & & & $14.3\pm1.1$ \\
Normalization \swift1$^{e}$ & &  $3.14 \pm 0.28 $ & & &  $3.46 \pm 0.13 $ \\
Normalization \swift2$^{e}$ & & & $ 3.0 \pm 0.4 $  & &  $ 3.16^{+0.13}_{-0.12}$  \\
Normalization \swift3$^{e}$ & & & & $3.83 _{ -0.17 }^{+ 0.24 }$ &  $3.98 \pm 0.16 $\\
alpha$^{f}$ & $ 0.71 _{ -0.04 }^{+0.07 }$ & & & & $ 0.437 \pm0.009 $\\
f$^{g}$ & $ 0.907\pm0.007$ & & & & $ 0.867 \pm0.002$\\
$ \chi^2$      & 485.585 & 159.38  & 102.54 & 124.19 &  973.65 \\
 DOF$^{h}$           & 476     &  128    & 98      &103     & 811 \\
$ \chi^2_\nu$  & 1.02    & 1.25    & 1.05    & 1.21   &  1.20 \\
\hline
\\
\underline {Model: Partial Covering}\\
$ N_H$$^{d}$ & $230^{+60}_{-50}$ & $88^{+55}_{-32}$ &185$^{+107}_{-69}$ &215$^{+230}_{-115}$ &$180\pm30$ \\
$\Gamma$$^{c}$ & 1.37$\pm0.03$ &1.24$\pm0.05$ &1.1$\pm0.05$ &1.31$^{+0.06}_{0.05}$ &1.28$\pm0.02$ \\
Normalization \chandra$^{e}$ & 7.0$^{+1.4}_{-0.5}$ & & & &10.8$^{+2.5}_{-1.9}$  \\

Normalization \swift1$^{e}$ &  & 3.25$\pm0.13$ & &  &3.39$\pm0.009$ \\
Normalization \swift2$^{e}$ & & &2.69$\pm0.13$ &  & 3.05$\pm0.009$  \\
Normalization \swift3$^{e}$ &  & & &$3.97\pm0.18$ & 3.89$\pm0.011$ \\

Normalization B \chandra$^{e}$ & 100$^{+100}_{-50}$  & & & &70$^{+30}_{-50}$ \\
Normalization B \swift1$^{e}$ &  &5$^{+7}_{-3}$ & & & 25$^{+10}_{-17}$ \\
Normalization B \swift2$^{e}$&  & &2.6$^{+1.7}_{0.9}$ &  & 50$^{+30}_{-20}$\\
Normalization B \swift3$^{e}$ &  & & &70$^{+140}_{-60}$ &36$^{+30}_{-16}$ \\

alpha$^{f}$ & 0.58$^{+0.06}_{0.07}$ & & & & 0.403$^{+0.040}_{-0.027}$ \\
f$^{g}$ & 0.85$\pm0.008$ & & & & 0.856 \\
$ \chi^2$ & 519.93 & 151.45 & 94.64 & 111.65 & 906.59  \\
DOF$^{h}$ & 475  & 127 & 97 & 102 & 807 \\
$ \chi^2_\nu$ & 1.09 & 1.19 & 0.98 & 1.09 & 1.12 \\
\hline 
\\
\underline {Fluxes and luminosities$^{i}$} \\
Flux$_{0.5-2\,keV}$ & 11.3$^{+1.3}_{-1.4}$       &  4.14$^{+0.12}_{-0.14}$&  3.59$^{+0.12}_{-0.11}$ & 4.96$\pm0.15$ &  & \\
Flux$_{2-10\,keV}$  &  33$\pm4$                  &  3.45$^{+0.07}_{-0.08}$&  8$^{+8}_{-5}$ & 4.0$\pm0.5$ &  & \\
Luminosity$_{0.5-2\,keV}$ & 7.7$^{+0.9}_{-1.0}$  &  5.30$^{+0.15}_{-0.18}$&  20.9$^{+0.7}_{-0.6}$ & 68$\pm2$  &  & \\
Luminosity$_{2-10\,keV}$  & 9.3$\pm1.1$          &  19$\pm0.4$&  100$^{+100}_{-60}$ & 320$\pm40$  &  & \\
\\
\hline \hline
\multicolumn{6}{l}{$^{a}$ Exposures are given after filtering all bad events. Values are given in ks. } \\
\multicolumn{6}{l}{$^{b}$ Count rates are given in counts/second.}\\
\multicolumn{6}{l}{$^{c}$ $\Gamma$ is the index of the power law.}\\
\multicolumn{6}{l}{$^{d}$ N$_H$ is in units of 10$^{22}\,cm^{-2}$.} \\
\multicolumn{6}{l}{$^{e}$ Normalizations of the power law model in units of $\rm{10^{-3}}$\,photons keV$^{-1}$ cm$^{-2}$ s$^{-1}$ at 1\,keV.}\\
\multicolumn{6}{l}{$^{f}$ alpha:parameter of {\it jdpileup} pile-up model. Represents the probability, per photon count greater than one, that the piled event is not rejected by the spacecraft software. }\\
\multicolumn{6}{l}{$^{g}$ f is a parameter of {\it jdpileup} pile-up model and represents the fraction of {\it pile-uped} events extracted.  }\\
\multicolumn{6}{l}{$^{h}$ DOF corresponds to the degrees of freedom.}\\
\multicolumn{6}{l}{$^{i}$ Flux in units of $\rm{10^{-12}}$~erg s$^{-1}$ cm$^{-2}$ and luminosity in units of $\rm{10^{43}}$~erg s$^{-1}$.}
\end{tabular}\\
$\,$
\end{table*}

\section{Discussion}
\label{dis}

The new optical spectroscopic data of \iras\, show it has complex emission line profiles (c.f., Fig. 4, 5, 6 and 7). For the narrow emission line profiles, the best model is obtained using two Gaussian components. Therefore, we find that \iras\, has a double narrow line component that is shifted from the systemic restframe by $\pm$1\,\AA. On the other hand, the best model for the broad Balmer emission line profiles was obtained with three Gaussian components. One for the blueshifted or MC component (FWHM $\sim$5043\,km\,s$^{-1}$), that shows no evidence of outflows or winds; a very broad component VBC (with a FWHM $\sim$7479\,km\,s$^{-1}$) that shows a redshifted asymmetry; and a third Gaussian component that was used to fit the red-peaklet or RP. 

The large \hb\ width and the modest FeII emission place \iras\ among Population B sources \citep{sulenticetal11}. Also, it is worth mentioning that the profile of the HeII$\lambda$4686 line is different from the profile of \hb. This result is in agreement with a VBC and also with having little or no contribution from the broad component as is frequently observed in Population B quasars \citep[e.g.,][]{1992ApJS...80..109B,2003ApJS..145..199M}.

A relativistic disk model was used to fit the H$\beta$ broad component without successful results. Therefore, one possibility is that the origin of the VBC is gravitational redshift due to the presence of a very close disturber \citep[e.g.,][]{2018FrASS...5...19B}. Also, the RP was found with a radial velocity shifted by $\sim$3500\,km\,s$^{-1}$, with respect to the systemic velocity. The analysis done to the 2D spectrum of \iras\, suggests that the RP has a projected separation of $\sim$33\,pc. 

The most striking peculiar feature in the spectrum is the RP. Its large radial velocity suggests that it might be a component of a (bound) double BLR associated with a second black hole orbiting around a central, more massive black hole. Therefore, the separation between the SMBH of \iras\, and the RP can be obtained (the projected separation of 33 pc is a broad upper limit) using the SMBH mass estimate of \iras\, and the radial velocity  of the RP of $\sim$3500\,km\,s$^{-1}$. The projected radial velocity $v_\mathrm{r} = v_{\mathrm{K}} \sin\theta\cos\phi$, where $v_{\mathrm{K}}$ is the Keplerian orbital velocity, $\theta$\ is the orbit inclination, and $\phi$\ an azimuthal angle ($ \phi = 0, \pi$\ if the velocity vector is aligned with the line of sight). Since the black hole mass is known, we can compute the orbital separation as 
$r ={G M_\mathrm{BH}}/{v_{\mathrm{K}}^{2}} \lesssim {G M_\mathrm{BH}}/{v_{\mathrm{r}}^{2}}$ and we obtain a separation $r\,\lesssim$\,0.1 pc. Based on this result, the RP and \iras\, could be forming a binary-AGN system with a sub-pc scale separation. On the other hand, the possibility that the RP is the result of outflows from circumnuclear star clusters seems unlikely because of both, the high velocity and the red-ward displacements found. Even in the case of it being due to super star clusters, expected outflow velocities are $\ll$ 1000\,km\,s$^{-1}$\ \citep[e.g.,][]{biketal18}, and in the case of luminous starbursts, outflow velocities are  $\lesssim 1000$ \kms \citep{letiranetal11}. Finally, the possibility of a feature associated with shocked gas in a hot spot within the accretion disk cannot be ruled out, and is actually a likely  alternative to a second BLR associated with a massive compact object (see the thorough discussion for two of the most-extensively-monitored type-1 AGN, NGC 4151 and NGC 5548 \citep[][]{bonetal12,bonetal16}). Therefore, in order to confirm the nature of the RP, spatially-resolved spectroscopic observations are needed.

Among the neighbours of \iras\, we find that apparently, there is a broad feature on the blue side of H$\alpha$ on \iras\, NE, shown in panel b of Figure~\ref{Fig:all-SPM}. This feature may not be a real broad component, but produced by the addition of narrow lines emitted by gas with a steep velocity gradient, as was found in  the study of the interacting mixed galaxy pair CPG 29 \citep[see][]{1994ApJ...435..668M}. The spectrum of 2MASXJ06021038+2828112, shown in panel c of Figure~\ref{Fig:all-SPM}, has stellar-like features. Finally, the spectrum of \twomas\, shown in panel d of Figure~\ref{Fig:all-SPM}, reveals that it is an obscured Sy2 galaxy.

The BPT diagrams of the bridge zone show that there is a LINER-like extended emission region. This is consistent with our results obtained in the 2D spectra analysis of \iras, where extended emission is found both in the blue and red spectral regions with an estimated size of $\sim$4.9\,kpc. This result suggest that the bridge of ionized gas is connecting both AGN probably as a consequence of an ongoing wet merger. For instance, the extended LINER-like emission has been proposed to be due to shocks \citep[e.g.][]{2011ApJ...734...87R}, so in this case, the shocked gas in the bridge zone will support evidence for the merger scenario.

The SMBH mass estimates for \iras\, were obtained applying two different corrections given in \citet{Marziani2019}. The results  are log\,M$\rm_{BHsymm}$\,=\,8.59\,$\pm$\,0.14 (M$_\odot$) and log\,M$\rm_{BHvir}$\,=\,8.44$\pm$0.13 (M$_\odot$) and both results are in agreement within a factor of $\sim$1.4. On the other hand, the SMBH of \twomas\, is log\,M$\rm_{BH}$\,=\,8.21$\pm$0.2 (M$_\odot$). This result was obtained using the M$_{BH}$-$\sigma_\star$ relation.

The \iras\, field of view has been pointed in X-ray three times with the \swift\, satellite, and once with \chandra, see Figure~\ref{fig:swiftchandra}, and it is included in the {\it XMM-Newton Slew-Catalog}. The exposures range from 2.4 to 14.5 ks. \iras\, is a bright source in X-ray with flux of the order of $\rm{3\times10^{-11}}$\,erg s$^{-1}$ cm$^{-2}$ and luminosity of the order of 10$\rm{^{44}}$\,erg s$^{-1}$ in the 0.5\,-\,10\,keV band, and has been detected in all observations (c.f. Figs. 12, 13 and 14). In fact, the \chandra\, observation suffers from pile-up. 

The \chandra\, spatial resolution allows detecting also the companion galaxy \twomas, embedded in the PSF of the \iras\, in the {\it XRT}-\swift\, observations. The location of both \iras\, and \twomas\, coincides with the location of the optical counterparts, as can be seen in Figure~\ref{fig:panschandra}. A closer look at the \chandra\, image (see Figure~\ref{fig:chandrasofthard}) shows that while \iras\, soft emission dominates over the hard emission, as expected for a Type 1 source, while \twomas\, soft emission is negligible, favouring a Type 2 class for this source. These findings are in good agreement with the results on the optical spectral analysis for both sources. 

The four available X-ray spectra of \iras\, have been extracted. All the spectra can be fitted with a single power law model. The inclusion of intrinsic absorption at the location of the galaxy does not improve any of the fits. The results of all fits can be seen in Table~\ref{Table:xrayfits}. Except for the 35255001 \swift\, observation, the fits are statistically acceptable, with $\rm{\chi^2_\nu}$ in the 1.22 to 1.00 range. However, this \swift\, image show hints of point-like X-ray sources surrounding the emission of \iras. The spectrum extracted in a region which avoids these sources, $r$\,=\,18\arcsec, provides a much better fit, $\chi^2_\nu$\,=\,0.99, but leaving unchanged the value of the power law index, $\rm{\Gamma}$\,=\,1.23$\pm$0.07. 

Source flux variations in \iras\, were detected. These are more evident in the soft band, and also in the power law index, being the \chandra\, ones showing the highest values. A comparison of the Chandra observed flux with the mean fluxes from the three SWIFT observations, in the soft and hard bands,
shows that Chandra flux is 2.6 and 1.6 higher, respectively.  The flux variation is evident by comparing these results with the flux estimated in the {\it XMM-Newton Slew Catalogue}, F$\rm{_{0.2-10\,keV}}$\,=\,1.1$\rm{\times}$10$^{-11}$\,erg\,s$^{-1}$\,cm$^{-2}$, and with the ones obtained in the three \swift\,and \chandra\, observations, in the same energy band: 3.2$\rm{\times10^{-11}}$\,erg\,s$^{-1}$\,cm$^{-2}$, 3.2$\rm{\times10^{-11}}$\,erg\,s$^{-1}$\,cm$^{-2}$, 3.5$\times10^{-11}$\,erg\,s$^{-1}$\,cm$^{-2}$, and 5.7$\rm{\times10^{-11}}$\,erg\,s$^{-1}$\,cm$^{-2}$, respectively. 
 
 The source is therefore clearly variable and although one of these fluxes is only an estimate, the flux of the \chandra\, observation, obtained in 2011 January 6 is more than five times higher than the lowest flux measured earlier by \textit{XMM-Newton} in 2003 March 26. 
The counts detected for \twomas\, in the \chandra\, observation are too low to allow an extraction and analysis of its spectrum. However, we have estimated the flux using the count rate of the source. The values obtained are
F$\rm{_{0.5-2\,keV}}$\,=\,6.3$\rm{\times10^{-16}}$\,erg\,s$^{-1}$\,cm$^{-2}$ and F$\rm{_{2-10\,keV}}$\,=\,5.0$\rm{\times10^{-14}}$\,erg\,s$^{-1}$\,cm$^{-2}$, being the soft band flux two orders of magnitude lower than that of the hard band, revealing the hard (obscured) nature of this source as a Type 2 AGN.

The multiwavelength analysis done in this work strongly suggests that the previously identify DAGN in \iras\, and \twomas\, consists of a pair of Sy1--Sy2 galaxies. The overlayed optical and radio maps presented in the lower panel of Figure \ref{Fig:VLA} show that \iras\, and \twomas\, are compact radio sources. The detection of \twomas\, in the radio bands, along with its black hole mass estimate and systemic velocity are presented for the first time in this work. Finally, the obtained projected separation with the VLA data of this DAGN system is 20.08\,\arcsec\, that corresponds to $\sim$13.3\,kpc.

\bigskip
\noindent{\bf Data Availability}
\bigskip

The optical data underlying this article will 
be shared on reasonable request to the corresponding author.
The PanSTARSS images are available at \url{ http://ps1images.stsci.edu/cgi-bin/ps1cutouts}.
The VLA data are available at \url{htpps://science.nrao.edu/vlass/data-access}.
Swift data are available at \url{https://swift.ac.uk/archive/} (Obs.ID 3525501,3525502,3525503). 
Chandra data are available at \url{https://asc.harvard.edu/cda/} (Obs.ID 12864).

\section*{Acknowledgements}
We thank the anonymous referee for useful comments and suggestions that helped to improve the manuscript. EB and ICG acknowledge support from DGAPA-UNAM grant IN113320. CAN thanks support from DGAPA-UNAM grant IN113719 and from CONACyT project Paradigmas y Controversias de la Ciencia 2022-320020. DRD acknowledges support from the Brazilian funding agency CAPES, via the PNPD program. JMRE acknowledges the Spanish State Research Agency under grant number AYA2017-84061-P. JMRE also acknowledges the Canarian Governement under the project PROID2021010077,  and is indebted to the Severo Ochoa Programme at the IAC. This work was partially supported by CONACYT research grant 280789. The \textit{WHT} is operated on the island of La Palma by the Isaac Newton Group in the Spanish Observatorio del Roque de los Muchachos of the Instituto de Astrof\'isica de Canarias. This work is partly based upon observations carried out at the Observatorio Astron\'omico Nacional on the Sierra San Pedro M\'artir (OAN-SPM), Baja California, Mexico.This work is partly based  upon observations collected at Copernico telescope (Asiago, Italy) of the INAF - Osservatorio Astronomico di Padova. Our thanks to the supporting staff during the observing runs. This research has made use of data obtained from the Chandra Data Archive and the Chandra Source Catalog, and software provided by the Chandra X-ray Center (CXC) in the application packages CIAO, ChIPS, and Sherpa. We acknowledge the use of public data from the Swift data archive. This research has made use of the NASA/IPAC Extragalactic Database (NED), which is operated by the Jet Propulsion Laboratory, California Institute of Technology, under contract with the National Aeronautics and Space Administration.

\bibliographystyle{mnras}
\bibliography{References} 
\bsp	
\label{lastpage}
\end{document}